\documentclass{article}
\usepackage[utf8]{inputenc}
\usepackage[pdftex]{graphicx}
\usepackage{amsfonts}
\usepackage{float}
\usepackage[mathscr]{eucal}
\usepackage{algorithm,algpseudocode}
\usepackage{amsmath}
\usepackage{xcolor}
\usepackage{hyperref}
\hypersetup{colorlinks,allcolors=black}
\usepackage{amsmath}
\newtheorem{definition}{Definition}
\newtheorem{assumption}{Assumption}
\newtheorem{axiom}{Axiom}
\newtheorem{theorem}{Theorem}
\usepackage{amssymb}
\usepackage{authblk}
\usepackage{pdfpages}
\usepackage{bbm}
\emergencystretch=\maxdimen
\title{Dynamic Function Market Maker (DFMM)}
\author{Arman Abgaryan, Utkarsh Sharma}
\affil{Supra DeFi Research}
\date{February, 2023}

\providecommand{\keywords}[1]
{
  \small	
  \textbf{\textit{Keywords---}} #1
}

\begin{document}

\maketitle

\setcounter{page}{1}

\begin{abstract}
    Decentralised automated market makers (AMMs) have attracted significant attention in recent years. In this paper, we propose an adaptive and fully automated Dynamic Function Market Maker (DFMM) that addresses the existing challenges in this space. Our novel DFMM protocol incorporates a compelling inventory risk management mechanism, comprised of a data aggregator and an order routing protocol. The proposed DFMM protocol operates on a data aggregation method that ensures maximum synchronisation with all price-sensitive market information, which has the effect of asserting the principle of one price and keeping markets maximally efficient. Our data aggregator includes a virtual order book, guaranteeing efficient asset pricing on local markets by ensuring DFMM price-volume dynamics remain synchronous with information from external venues, including competitors. This data aggregation capability is a fundamental feature of the protocol and provides critical information for a novel rebalancing and order routing method. The rebalancing and order routing method in DFMM optimises outstanding inventory risk through arbitrageurs, in such a way that arbitrageurs are more likely to act to help DFMM manage its inventory, than another protocol without the aforementioned advantages, which is complemented with a price assurance mechanism that are primitive to the protocol. To manage risk, DFMM incorporates built-in protective buffers using non-linear derivative financial instruments. These buffers enhance the stability of the protocol and mitigate potential losses caused by market volatility. Additionally, the protocol employs an algorithmic accounting-asset, serving as the single asset that connects all the pools, and resolves the issue of segregated pools and throttled risk transfer. The settlement process is entirely protocol-driven, maximising the efficiency of risk management processes, and eliminating subjective market risk assessments. In essence, the proposed DFMM protocol offers a fully automated, decentralised, and robust solution for automated market making. By addressing inventory risk management through data aggregation, rebalancing strategies, and risk transfer mechanisms, DFMM aims to provide long-term viability and stability in an asset class that demands robustness.
\end{abstract}

\keywords{Dynamic Function Market Maker, automated market maker, decentralised finance, inventory risk management, data aggregation, order routing, risk transfer, non-linear financial instruments, decentralised exchange aggregation.}

\section{Introduction}
    A blockchain-based automated market maker (AMM) revolutionises trading by facilitating the exchange of value among market participants through smart contracts, eliminating the need for a centralised matching agent. These systems have rapidly gained popularity, supporting significant volumes of financial assets. Their appeal stems from not relying on active market makers but instead leveraging liquidity providers (LPs), as well as their inherent robustness compared to vulnerable ``centralised" exchanges prone to adversarial attacks and malicious operators. Despite their success, there are fundamental deficiencies (e.g., \cite{park2021conceptual, angeris2020improved}) that hinder wider adoption, including high slippage, impermanent loss, and unsatisfactory risk mitigation protocols.\\
    \\
    To address these limitations, we introduce the Dynamic Function Market Maker (DFMM), a blockchain-based AMM with dynamic functionalities designed to overcome shortcomings in popular AMM designs. First and foremost, DFMM enables the aggregation of price-relevant information from external trading venues, whether centralised or decentralised, through a virtual order book and decentralised oracle service. This mitigates the impact of impermanent loss arising from disconnected markets operating in isolation, reducing the reliance on arbitrageurs to synchronise AMM prices with prevailing prices in competing venues. Second, DFMM optimises liquidity distribution across different price levels, effectively minimising slippage. Third, DFMM ensures maximal stability by actively tracking emerging market risks through a rules-based mechanism. This informs the management of outstanding inventory risks, supported by the introduction of secondary liquidity providers (sLPs), new agents responsible for pricing and trading inventory risk. The system prices the expediency of risk management through an auction-based mechanism that determines rebalancing premia. Lastly, we introduce the concept of an algorithmically managed accounting asset, an accounting asset acting as a common counterpart for all pools. This facilitates liquidity aggregation across disjointed asset-pair pools and enables optimised liquidity allocation, concentrating liquidity near the current market price leading to an optimal price-volume curve.\\
    \\
    A schematic of the DFMM framework can be found herewith.
    
    \begin{figure}[H]
      \begin{center}
      \includegraphics[width=4.2in]{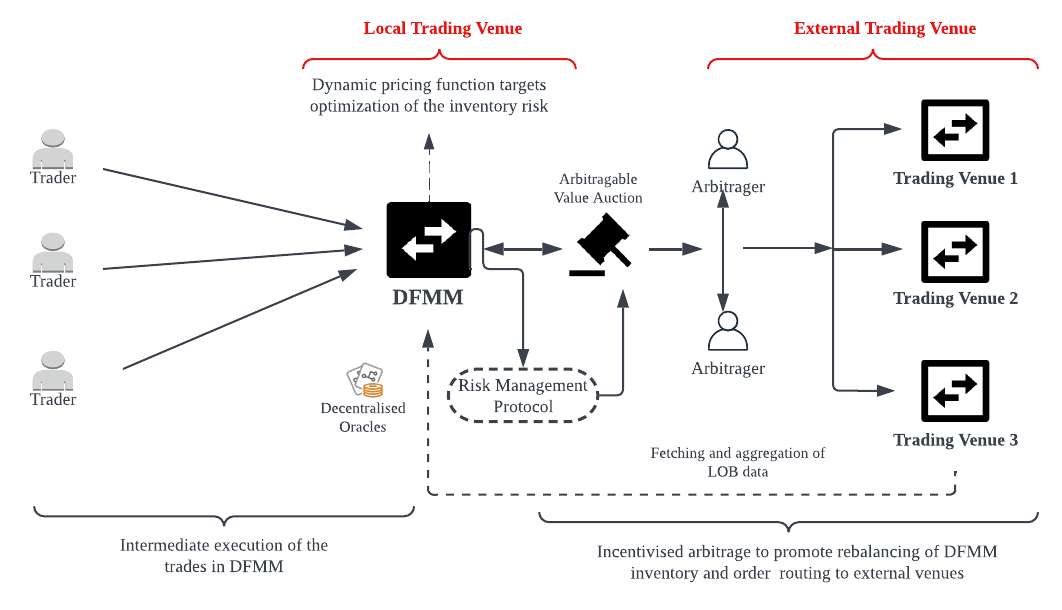}
      %\vspace{-15pt}
      \caption{DFMM schematics.}\label{dfmmschematics}
      \end{center}
    \end{figure}
    
    \noindent
    DFMM is a robust and comprehensive solution that serves as a one-stop-shop for trading digital financial assets, offering multi-exchange and cross-chain liquidity aggregation. Its advantages over single-venue-based price discovery are evident, as it effectively limits slippage, mitigates impermanent loss, and significantly reduces inventory risks. By promoting fundamental value discovery, DFMM contributes to the progression of the digital assets industry, ensuring that the sustainability of value is not solely dependent on the inflow of new capital. Additionally, DFMM stands out as a superior competitor to decentralised exchange (DEX) aggregators. It excels in three key aspects: (i) providing financial incentives for routers to optimise the efficiency of order execution paths, rather than solely focusing on maximising the executed volume; (ii) eliminating barriers to entry for order routing service providers, by enabling agents such as miners, who are well-positioned to act as routers, to serve as routing service providers on the network; and (iii) fully decentralises the exchange aggregation and order routing mechanisms, removing the need for centralised operators.\\
    \\
    The remainder of this paper is organised as follows: in Section 2, we cover preliminary concepts, and outline our assumptions and definitions. Section 3 provides a summary of existing works in the field. In Section 4, we provide a comprehensive overview of the DFMM protocol, with key subsections covering the external market curve, liquidity provision, local price formation, dynamic regulation, and protocol P\&L. In the forthcoming revisions, we will further elaborate on the framework, and present results of our simulation, including stress tests, offering valuable insights into the performance and effectiveness of DFMM.

\newpage
\section{Preliminary Concepts \& Assumptions}\label{prelims}
   
    \begin{assumption}
        Market participants in the internal and competing external market are rational, which we define using utility maximisation\cite{mccormick1997essay}.\\
        \\
        Let's consider an individual with a set of alternatives, denoted by $A$, and a preference relation over these alternatives, denoted by $\succcurlyeq$. The individual's preferences satisfy the following axioms:
        \begin{enumerate}
            \item Completeness: For any two alternatives $x$ and $y$ in $A$, either $x\succcurlyeq y$ (alternative $x$ is preferred to or indifferent to $y$) or $y\succcurlyeq x$ (alternative $y$ is preferred to or indifferent to $x$).
            \item Transitivity: If $x\succcurlyeq y$ and $y\succcurlyeq z$, then $x\succcurlyeq z$. In other words, if alternative $x$ is preferred to or indifferent to $y$, and $y$ is preferred to or indifferent to $z$, then $x$ is preferred to or indifferent to $z$.
            \item Continuity: For any three alternatives $x$, $y$, and $z$ in $A$, if $x\succcurlyeq y\succcurlyeq z$, then there exists a positive number $\alpha$ such that $y\succcurlyeq \alpha x + (1-\alpha)z$. This axiom ensures that small changes in the attributes of alternatives result in small changes in individual preferences.
        \end{enumerate}
        
        \noindent
        Given these axioms, an individual is considered rational if they maximise their utility function. The utility function $U: A \rightarrow \mathbb{R}$ assigns a real number to each alternative in $A$, representing the individual's subjective satisfaction or preference for that alternative. Mathematically, a rational individual chooses the alternative $x$ that maximises their utility function:
        \begin{equation}
            x \in A: U(x) \geq U(y) \ \forall y \in A.
        \end{equation}
    \end{assumption} 

    \noindent
    Furthermore, it is the rationality of market participants, that motivates them to be solely interested in maximising their terminal wealth $W_{T}, \forall T \in (0,\infty)$, where the initial condition satisfied is $W_0 > 0$. We assume that LPs are interested in maximising wealth over a longer horizon than other market participants, like short-term traders.
    
    \begin{assumption}
        Future asset price outcomes in the indigenous and competing external market are unbounded, belonging to the positive real number space, denoted as $\mathbb{R}^{+}$. Mathematically, it can be expressed as $\lim_{t \to \infty} P(t) = \infty$, where $P(t)$ represents the asset price at time $t$. However, due to the computational intractability of analysing an infinite set of potential outcomes, it is practically infeasible to exhaustively compute the complete set of possibilities.
    \end{assumption}
    
    \begin{assumption}
        Considering an asset $X$, the volume traded on external venues ($V^{E_X}$) is significantly greater than the volume traded in our internal market ($V^{I_X}$), i.e., $V^{I_X} < V^{E_X}$. This observation leads us to rely on the rationality of market participants and assert that it is the local market price that would need to adjust to accommodate dynamics evolving in external markets, rather than the reverse.
    \end{assumption}
    
    \begin{assumption}
        The DFMM system evolves over time with discrete time steps denoted by $t$. The time step is defined based on the sequential state updates in the protocol governing the system's dynamics. It is important to note that different time steps within the system do not necessarily have the same duration.
    \end{assumption}

\begin{definition}[Aggregators]
    Aggregators are specialised entities that consolidate the trading volume of digital assets by pooling together liquidity from various trading venues or platforms, like decentralised and centralised exchanges. By leveraging advanced technology and algorithms, they help to create a unified and streamlined marketplace, offering traders access to a larger and more diverse pool of liquidity. This allows for improved trade execution, reduced slippage, and increased overall efficiency in the digital asset trading ecosystem.
\end{definition}
    
    \begin{definition}[Epoch]
        We define an epoch($e$) in terms of a fixed time interval, e.g. block time, such that a single epoch can include multiple time steps.
    \end{definition}
    
    \begin{definition}[Automated Market Maker]
        An automated market maker (AMM) is a financial protocol or smart contract-based entity that operates as a decentralised financial institution. It utilises a pre-set rules-based mechanism, often relying on a deterministic payoff function, to control the pricing ($P^X_{t}$ and $P^Y_{t}$) and trading volume ($V^X_{t}$ and $V^Y_{t}$) of specific digital financial assets within the market (in this case, for assets $X$ and $Y$.).
    \end{definition}
    
    \begin{definition}[Cross-chain Automated Market Maker]
        A cross-chain Automated Market Maker (cAMM) is a blockchain-based trading system that operates across multiple blockchain networks. It can be represented as a dynamic function $AMM: \{\mathcal{B}_1 \ldots \mathcal{B}_n \} \rightarrow \mathcal{T}$, where $\mathcal{B}_i$ represents the blockchain network $i$ and $\mathcal{T}$ denotes the trading functionality.\\
        \\
        The cAMM enables the trading of digital financial assets hosted on different blockchains, denoted as $A_1, A_2, \ldots, A_m$, where each $A_i$ belongs to its respective blockchain network $\mathcal{B}_i$. The trading functionality allows users to exchange these assets, potentially involving swaps, liquidity provision, or other trading mechanisms across the different blockchain networks.
    \end{definition}

    \begin{definition}[Inventory Risk Management]
        Inventory risk management aims to minimise potential losses resulting from imbalances in the market participant's inventory. If $\mathcal{I}$ represent the inventory of assets or positions held by a market participant, and $\mathcal{P}$ represent the reduced positional exposure after risk management, then it can be formulated as an optimisation problem, 
        \begin{equation}
             Min\text{Loss}(\mathcal{P}),
        \end{equation}

        \noindent
        where $\mathcal{P} \text{satisfies risk tolerance constraints}$.\\
        \\
        The practice of inventory risk management involves actively managing the risk associated with holding assets or positions, aiming to reduce overall positional exposure and minimise potential losses resulting from imbalances in the market participant's inventory.        
    \end{definition}

\begin{definition} [Impermanent Loss]
    Impermanent loss (IL) refers to a specific type of market risk inherent in Automated Market Makers (AMMs) that utilise an arbitrageur-driven price discovery mechanism, such as Constant Product Market Makers (CPMMs). In a constant product environment, changes in the external market price of an asset incentivise trading by arbitragers to adjust inventory levels in AMM to remove the price discrepancy between the AMM and external marketplaces. As a result, the arbitrager profits at the expense of the liquidity provider. In the constant product market maker with two assets($X$ and $Y$) in the liquidity pool, the impermanent loss after price change can be expressed as follows:  
    \begin{equation}
        IL = \sqrt{\frac{P_{1}^X}{P_0^X}} - \frac{\frac{P^X_{1}}{P^X_{0}}+1}{2},
    \end{equation}
    \noindent
    where,\\
    $IL$ represents the impermanent loss,\\
    $P_0^X$ denotes the initial external market price of the asset $X$  in terms of asset $Y$ at the point of provision of liquidity to the pool by LP,\\
    $P_{1}^X$ denotes the subsequent external market price of the asset $X$ in the terms of asset $Y$.
\end{definition}

\begin{definition} [Liquidity]
    Liquidity is a catch-all term used to describe the ease with which an asset can be bought, or sold, in the market, without significantly impacting it's price. There are several ways to measure this, but in this work, we do it by cumulating volume available to buy or sell an asset, at different price points.\\
    \\
    Let $\mathcal{L}(P_L, P_U)$ represent the liquidity between two price points $P_L$ and $P_U$, which is the cumulative volume available in the order book between these two price points:
    \begin{equation}
        \mathcal{L}(P_L, P_U) = \int_{P_L}^{P_U}V(p)dp.
    \end{equation}
    
    \noindent
    Visually, this can be represented on a price-volume chart (liquidity density function\footnote{Liquidity density function is a function that provides the density of volumes available at different price. Readers would observe that this is closely linked to the liquidity vector, defined early on in the section, such that two liquidity vectors for bid and ask side can be said to form an orderbook. However note, that we transform this function, to better fit DFMM.}), where the x-axis represents the price and the y-axis represents the depth (or volume) available at each price level. If the mid-market price is at the centre of the x-axis, then the left-hand side of the axis represents the price-volume dynamics for bids, and the right-hand side of the axis represents the price-volume dynamics for offers.
\end{definition}

\begin{definition} [Concentration Rate of Liquidity]
    The concentration rate of liquidity, denoted as $\text{CR}$, is a measure indicating the proximity of the allocation of liquidity around the current market price.
    \begin{equation}
        \text{CR} = \frac{\sum_{i=1}^{n} \mathcal{L}_i}{\sum_{i=1}^{n} \mathcal{L}_i \cdot |P_i - P_{\text{market}}|},
    \end{equation}

    \noindent
    where,\\
    $\text{CR}$ represents the concentration rate of liquidity,\\
    $\mathcal{L}_i$ represents the liquidity available at price level $i$,\\
    $P_i$ represents the price level,\\
    $P_{\text{market}}$ represents the current market price.\\
    \\
    The concentration rate of liquidity is calculated by summing the product of liquidity ($\mathcal{L}_i$), and the absolute difference between each price level ($P_i$) and the current market price, say mid-price, ($P_{\text{market}}$), divided by the total liquidity available, across all price levels. This calculation quantifies the degree of liquidity concentration near the current market price. A higher concentration rate indicates that a significant portion of liquidity is concentrated closer to the current market price, while a lower concentration rate suggests a more dispersed allocation of liquidity across various price levels.
\end{definition}

\begin{definition}[Slippage]\label{slippage_definition}
    Slippage can be defined at time $t=t$,the difference between the expected price ($P^*_t$) and the actual executed price ($P_t$),
    \begin{equation}
        S_t = P^{*}_t - P_t.
    \end{equation}
    
    \noindent
    Note, that the expected price could be, say, the best bid or offer, or any other price which the executing agent believes (or hopes) to get their target volume executed at. Further note, that slippage may also sometimes result in a positive impact on the trader's P\&L, when the stochasticity in the price of execution turns out to be favourable.\\
    \\
    Some of the factors that impact slippage include order-book dynamics, i.e. the bid-ask spread, volumes available for execution, and volume the trader is seeking to execute in the trade, which is often captured in market impact literature (e.g. see \cite{bacry2015market}).
\end{definition}

\begin{definition}[Market Impact]\label{marketimpact_definition}
    Market impact refers to the impact of a trade on the price of an asset.
    \begin{equation}
        \Delta P_t = \alpha V_t, 
    \end{equation}

    \noindent
    where, $\Delta P_t = P_t - P_{t-1}$, $\alpha$ is a coefficient capturing the sensitivity of the price to trade size, and $V_t$ is the size of the trade\footnote{Several competing models (e.g. linear, quadratic, power-law, etc.) of the aforementioned exist in financial literature, but for the purposes of this work, the aforementioned formulation suffices.}.
\end{definition}

\begin{definition}[Liquidity Pool]
    A liquidity pool is a crowd-sourced pool of digital assets locked in a smart contract that is used to facilitate trades in the market. We denote the liquidity pool (smart contract) holding asset $X$ by $L^X$.  Furthermore, by referring to liquidity pools as paired we assume that assets held in these pools can be swapped with each other. We express the paired liquidity pools (or just pair) of asset $X$ and $Y$  in terms of tuple $\{L^X, L^Y\}$.  The traders that put the asset in one of the paired pools (such as $L^X$), receive the asset from the paired pools (such as $L^Y$), where the amount of the asset received is based on the market pricing rule.
\end{definition}

\begin{definition}[Liquidity Vector]
    A liquidity vector incorporates the volume available at different price points, $\mathbf{\mathcal{L}} = (V_{P_L}, \dots, V_{P_H})$, with the number of elements equal to the $\frac{P_{H} - P_{L}}{\delta P}$, where $\delta P$ is the minimum price-step, and $L$ and $U$ representing lower and upper values. This allows a participant to compute the total volume of an asset - indicated by the superscript, which can be purchased with a limit price $P = P_H$,
    \begin{equation}
        V_{P_H} = \int_{0}^{P_H} \mathbf{\mathcal{L}}(P) dP.
    \end{equation}
\end{definition}

\noindent
Note, that in the preceding three definitions, we have focused on the price dimension to link volumes available for any asset. We do this for ease of our readers' convenience, as it directly links to the manner in which limit order books are typically represented. However, as it will become clear in the sections that follow, it is mathematically more convenient for us to use volume dimension as the focal point of analysis.\\
\\
We now list key stakeholders of AMMs, which are as follows \cite{heimbach2021behavior, evans2020liquidity}:

\begin{definition}[Market Makers]
    Market makers are financial market participants providing prices at which they are willing to not just buy (or sell), but also sell (or buy) any product. Typically, this is done within a strict risk management framework, which often requires running as small a net position as possible, by adapting the bid or ask price accordingly. Since market makers take the risk of buying when everyone is selling, or selling when everyone is buying, they charge a theoretical fee on each transaction, which is the bid-ask spread, which in the context of DeFi AMMs is often in terms of the AMM fees. However, this fee may not sufficiently compensate for the risks arising from market-making activities. Among others, this is largely due to inventory risk stemming from imbalanced positions and the risk that the price of an asset will move significantly from the current market price due to market pricing rules (inherent in some AMMs, e.g. CFMM) leading to a deviation beyond the current fair value(impermanent loss).
\end{definition}

 \begin{definition}[Primary Liquidity Providers]
    Primary liquidity providers (pLPs) are market participants who aim to maximise their terminal wealth, denoted as $W_T$ for all valid time steps $T$ within the range $(0,\infty)$. It is assumed that pLPs start with an initial wealth value of $W_0$, which is greater than zero. pLPs submit digital assets to DFMM, and by doing so, they provide a stable source of liquidity. Unlike risk-seeking market participants who actively optimise their risk positions, pLPs are primarily passive risk-takers, and hence, in comparison to other agents described in this paper, are the slowest to rebalance their inventory to a new signal. Their focus is on earning an income by providing liquidity, rather than actively seeking risk or maximising returns. The total amount of inventory of a specific asset, denoted as $X$, that pLPs have submitted at time step $t$, is represented by $\mathcal{I}^X_{{LP}_t}$. This quantity captures the inventory held by pLPs and made available for trading in the AMM.
\end{definition}

\begin{definition}[Arbitrageurs]
    Arbitrageurs help enforce the law of one price (e.g. see:\cite{froot1995law}) of the same asset across different venues, by exploiting informational inefficiencies which lead to price asynchronicity. Their role as a financial market participant is particularly important for blockchain-based AMMs due to the inherent inability of the blockchain-based systems to interact with external systems, i.e. systems residing outside of the specific blockchain infrastructure. These participants risk their own capital, seeking to maximise their wealth at the next time step, i.e. $W_{t+1}$ and not $W_T$ where $T >> t$. The expected reward is represented by the total arbitrageable value in the system. Therefore, unlike pLPs, arbitrageurs are simply interested in maximising their wealth at the next time step.\\
    \\
    An arbitrageur searches all AMM pools of digital assets for investable opportunities yielding the maximum encashable arbitrageable value, and compares them against its presumed fixed costs, which include the opportunity cost of capital, any gas fees payable on AMM or other systems, and other expenses. We assume no market impact, resulting from an arbitrageur's decision to act, and focus on the rebalancing effect of their activities in the pool.\\
    \\
    Finally note, that there is an operational risk that each arbitrageur is exposed to, resulting from the multi-legged transaction they have to engage in, in order to fully monetise the arbitrageable value associated with any digital asset pool. This operational risk is often limited by setting an upper limit on the maximum exposure an arbitrageur is willing to carry, which is meaningfully smaller in comparison to a trader seeking to build positions for long-term holding.
    
\end{definition}

\begin{definition}[Traders/Liquidity Takers]
    Traders (or liquidity takers) are market participants who actively engage in buying and selling securities or assets with the objective of building medium to long-term positions. They contribute to the stability and growth of the asset's market by providing long-term demand, and strategically investing in the digital finance asset space, anticipating potential value appreciation or depreciation. They aim to profit from price movements away from value, and exploit market inefficiencies over an extended period. Traders are seen as end-users of liquidity and build substantial positions, therefore, prioritise minimisation of their trade's market impact, and slippage. In some of the popular AMMs, traders define a slippage tolerance parameter, representing the maximum acceptable deviation from the target price.\\
    % They execute trades up to the tolerance level, halting trade execution once the limit is reached.\\
    \\    
    More formally, let's denote:
    \begin{itemize}
        \item $T$: The set of traders.
        \item $S$: The set of securities or assets.
        \item $t \in T$: A specific trader.
        \item $s \in S$: An asset.
        \item $P(t, s)$: The price of security $s$ as perceived by trader $t$, which could be the best price that can be traded in the entire market.
        \item $V(t, s)$: The trader $t$'s volume or position in security $s$.
        \item $\text{Slippage}$: The slippage, as defined in Def. \ref{slippage_definition}.
        \item $\text{MarketImpact}$: The market impact parameter, representing the impact of trader $t$'s transactions on the market.
    \end{itemize}

    \noindent
    The objective of the trader $t$ is to build a long-term position while minimising (negative i.e. loss-making) slippage and market impact. The trader aims to accumulate liquidity and minimise the deviation from the target price during the execution of trades. To represent this objective mathematically, we can use the following formulation:
    \begin{align*}
        \text{minimise:} \quad & \sum_{s \in S} \sum_{t \in T} (\text{Slippage}(t, s) \cdot V(t, s)) + \text{MarketImpact}(t, s) \\
        \text{s.t.:} \quad & V(t, s) \geq 0 \quad \forall t \in T, s \in S \quad \text{(Non-negativity constraint on trader positions)} \\
        & \text{MarketImpact}(t, s) = f(V(t, s)) \quad \forall t \in T, s \in S \quad \text{(Market impact as a function of trader's volume)}
    \end{align*}

    \noindent
    Here, the first objective term represents the accumulated slippage cost for all traders and securities, while the second term represents the total market impact caused by the traders' transactions\footnote{Specific mathematical forms of $\text{Slippage}(.)$, $\text{MarketImpact}(.)$, and the relationship between $\text{MarketImpact}(.)$ and trader's volume (function $f$) would need to be defined based on the specific modelling assumptions and considerations of the trading environment.}. The constraints enforce the non-negativity of positions, the slippage tolerance, and the relationship between market impact and trader volume.
\end{definition}

\newpage
\section{Literature Review}
    Trading of products, whether financial or otherwise, has a long history rooted in centralised markets. As financialisation and increased centralisation took hold, the role of market makers emerged to maintain order in these markets. Traditionally, regulated financial entities have been entrusted with this function, acting as principal risk takers who usually seek to profit from the bid-ask spread, at minimal trading fees, whilst aiming to minimise inventory risk. In return, these market makers are obligated by the exchange on which they operate to provide quotes of a predetermined quality. Failure to fulfil these market-making obligations can result in fines or the loss of privileges, such as preferential fee structures.\\
    % This model has proven successful in traditional financial markets, particularly when the underlying products are supported by tangible contracts like stocks, warrants, or bonds.
    \\
    As the evolution of digital assets gathered pace, it became clear that traditional models of operating risk in markets have deficiencies for use, especially as instances of fraud, such as those witnessed in FTX and Todex, as well as infrastructural deficiencies in exchanges like Bitfinex and Mt. Gox, have highlighted the risks and vulnerabilities of centralised systems. This shouldn't be a surprise, as it is a desire to operate a trustless system, which is at the heart of innovation in these digital assets, so it is anyway sub-optimal that these systems are operated in a centralised manner, leaning on trust on a single or a meaningfully concentrated group of entity/entities. This has refocused attention on decentralisation, with an emphasis on distributed networks and peer-to-peer transactions, presented as an alternative paradigm for trading and asset exchange. It offers the potential for increased transparency, reduced counterparty risk, and enhanced user control over assets. This has naturally led to the need for intermediaries like traditional market makers and centralised exchanges to be re-evaluated.\\
    \\
    These issues are complemented by computational limitations which have hampered the implementation of centralised limit orderbook style systems, which has led to the innovation of automated market makers \cite{wang2020automated, mohan2022automated, niemerg2020yieldspace, slamka2012prediction} and decentralised exchanges. Albeit still maturing, these decentralised and automated mechanisms have since witnessed significant advancements and iterations. These automated rules-based mechanisms are driven by mathematical logic, for example, see \cite{jensen2021homogenous}, which are meaningfully limiting in their ability to offer a comprehensively dominant alternative to the state-of-the-art implementations in the traditional finance world. Some market makers are driven by limited(constant) mathematical functions\cite{angeris2020does, angeris2020improved, angeris2021constant}, leading to significant inefficiencies for agents utilising them. Some of the more noteworthy works include, Constant Product Market Maker (CPMM)\cite{adams2021uniswap, bworld}, Constant Mean Market Maker (CMMM)\cite{Balancer}, and Constant Ellipse Curve Market Maker\cite{ wang2020automated}. It might also be worth noting that well before the popularisation of blockchain technology, several automated market-making frameworks existed, e.g. Combinatorial Information Market\cite{hanson2003combinatorial}, modular combinatorial information aggregator using logarithmic market scoring rules \cite{hanson2007logarithmic}. Arguably step-up from the earlier, iterations is the paradigm of hybrid market makers, which combines different types of pricing functions for stablecoins, e.g. \cite{curve}, liquidity provisioned through a constant power sum formula \cite{niemerg2020yieldspace}, a weighted constant product market maker focusing on the issue of fragmented liquidity  \cite{curve} inspired hybrid market maker focusing on perfecting market impact.\\
    \\
    % \cite{ramabaja2023xfai},\cite{kim2023fintech}  \cite{cantarutti2023silkswap}
    However, whilst decentralised trading and market-making offer exciting possibilities, challenges remain. Issues such as competitive price, scalability, regulatory frameworks, and user experience need to be addressed for widespread adoption.\\
    \\
    The dynamic function market maker we propose in this work, is an AMM, with price-volume curve having the dual functionality of dynamic curves \cite{krishnamachari2021dynamic} and order routing - for e.g. used in \cite{angeris2022optimal, danos2021global}. There has been some academic research exploring the optimal execution of the trade routing in DeFi, but to the best of our knowledge, there are no similarly well-adopted protocol layer propositions for the decentralised incentivisation of optimal routing. Related to these order routing mechanisms, is the DEX aggregation and routing protocols (e.g. see 1inch\cite{1inch}), Paraswap\cite{Paraswap}, and OpenOcean\cite{openocean}. A further step from the order routing paradigm, is the one involving a network of AMMs\cite{engel2021composing}. Finally, one relatively unexplored area in this field is that of optimal risk management processes for automated market makers, which we emphasise through our simulations.\\
    \\
    The use of leverage in these systems is another area of exploration, e.g. in \cite{wang2022speculative}, authors deploy leverage in an under-collateralised environment and suggest solutions to the risk of impermanent loss, arbitrage loss, and collateral liquidation. However, it appears that instead of solving causal factors at play, these works try to deploy a make-shift solution, and present it as one that's sustainable in the long term.

\newpage
\section{The Protocol}
    The Dynamic Function Market Maker (DFMM) is a system that seeks to provide efficient market-making services while adapting to changing market conditions. The proposed mechanism denoted as $\Omega(.)$, consists of three functions: $\mathcal{L}^I(.)$, $\mathcal{L}^E(.)$, and $\mathcal{R}(.)$. These functions utilise internal information ($\omega^I$), and external information ($\omega^E$) to facilitate market-making operations. The external information, represented by $\omega^E$, is utilised to infer an external liquidity density function denoted as $\mathbf{\mathcal{L}^E(.)}$. This inference process is described in detail in Section \ref{externalcurve}. The function $\mathcal{L}^I(\mathcal{L}^E(\omega^E), \mathcal{R}(\omega^I))$ represents the internal liquidity density function, which is a function that maps the combined external liquidity information $\mathcal{L}^E(\omega^E)$ and the internal rebalancing information $\mathcal{R}(\omega^I)$ to the internal liquidity density function. One of the distinguishing features of the DFMM system is the rebalancing function ($\mathcal{R}(.)$), which plays a crucial role in effectively distributing liquidity across different price levels and facilitating trade routing to competing external markets.\\
    \\
    The remainder of this section is organised as follows: in section \ref{externalcurve}, we describe how prices from external venues are used to construct a virtual order book reflecting external liquidity density function; in section \ref{liquidityprov} we describe the algorithmic accounting asset and the price assurance mechanism; in section \ref{localprice} we describe how the protocol price if formed and updated, incorporating information from the external market; in \ref{arbvalue} we discuss the expedience with which the protocol seeks to augment its prices with information gleaned from external sources; in section \ref{protocolpnl} we discuss how the protocol's reserves are used in pursuit of protocol's goals, and the desirable behaviours it seeks from agents in the ecosystem.

\subsection{External Liquidity Density Function (ELDF)}\label{externalcurve}
    In section \ref{prelims} we assumed that the market is always right, which in the case of our DFMM implies that evolving price-relevant dynamics in external (digital asset markets other than DFMM) venues, must be reflected in the local venue. Thus at its simplest, i.e. in the linear form, the relationship between the volume of two assets on the external venue could be modelled as $V^X = \mathcal{L}^{E}_t (V^Y)$, where $V^X$ and $V^Y$ represents the volume of one asset which can be bought using the other asset, and $\mathcal{L}^{E}_t(.)$ represents the function mapping volume of one asset to equivalent volume in another asset at time $t$, using price information from external trading venues. We seek to learn this function - in our case, the external liquidity density function, using a second-order polynomial, and ascertain its coefficients using decentralised oracles.\\
    \\
    This allows us to learn about competing venues, and maximally minimise the risk of front running, among other punitive costs stemming from pricing inefficiencies. Note, it is the price difference between the DFMM and competing venues, which arbitrageurs seek to monetise, and thus, their payoff function is $\mathbb{E}[R_t^{arb}] = |{P^I - P^E}| (1 - \upsilon)$, where $\upsilon$ is the source of stochasticity in an arbitrageur's returns.

\subsubsection{ELDF Construction}
    To construct the ELDF curve, a decentralised data oracle gathers data points $(p_1, v_1), (p_2, v_2), \dots, (p_n, v_n)$ representing the available volume at each price point (denominated in, say, USD). In continuous trading, it is assumed that crossed markets cannot exist. Hence, all price levels are aggregated, as it is not crucial whether they represent bid or ask prices. These aggregated data points are then utilised to derive the external bid and ask liquidity density functions at time-step $t$, denoted as $\mathcal{L}^{E_X}_{\text{bid}_t}$ and $\mathcal{L}^{E_X}_{\text{ask}_t}$, respectively.
    
    \begin{figure}[H]
        \begin{center}
            \includegraphics[width=3.8in]{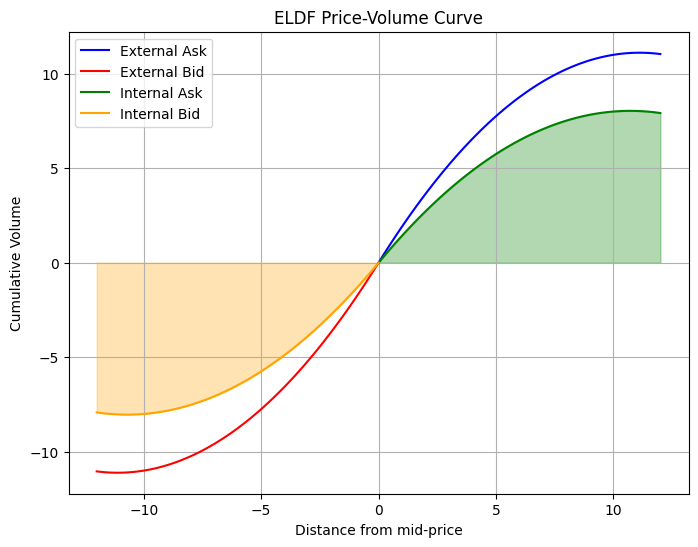}
            \caption{Example bid and ask curves.}
            \label{fig:eldfcurves}
        \end{center}
    \end{figure}

    \noindent
    In Fig.\ref{fig:eldfcurves}, we show how additional liquidity might be available to buy (or sell) an asset might be available in external markets, which can be quantified using the area between the internal and external curves for the ask (first quadrant) and bid (third quadrant)\footnote{Negative volume implies, the volume where the market does not have an inventory but is seeking to build the reflected amount by submitting bids, at prices lower than the mid-price point.}\footnote{While the plot shows that internal and external bid and ask curves never intersect, there might be scenarios where even though the external market has greater overall volume available, the internal market has a higher concentration of volume near the mid-price point.}. However, given some of the limitations of the currently available oracle-based solutions, it may be more practicable to combine bid and ask curves, into a single price-volume curve for a particular asset. These data points are used to fit a curve that provides a continuous representation of the market's order book. This enables the DFMM pricer to adapt its parameters and minimise the pricing discrepancies between internal and external markets, thus safeguarding stakeholder interests. In general, given $m$ pairs of data points $(x_i, y_i)$, where $i=0,1,\dots,m$, it is possible to fit an n-th order polynomial represented as:
    \begin{equation}
        y_i = f_n(x) = c_0 + c_1(x-x_0) + c_2 (x-x_0)(x-x_1) + \dots + c_n(x-x_0)(x-x_1)\dots(x-x_m),
    \end{equation}
    
    \noindent
    Alternatively, this polynomial can be represented more concisely as:
    \begin{equation}
        y_i = f_n(x) = \sum_{i=0}^{m}c_i\prod_{j=0}^{i-1}(x-x_j).
    \end{equation}
    
    \noindent
    This curve fitting process, i.e. finding the coefficients $c_0, c_1, \dots, c_n$, allows for the continuous representation of the CLOB, providing a foundation for the DFMM system to adapt to market dynamics and maintain alignment between local and external markets. The resulting bid and ask curves aid in accurate price determination and liquidity provisioning.\\
    \\
    The polynomial we obtain after solving, $f_n(x_m, x_{m-1}, \dots, x_0)$, where a polynomial function of a pre-specified order\footnote{We expect a second order polynomial to reasonably capture the dynamics we wish to focus on.} is the fitted market curve expected to help us quantify the manner in which our own local curve differs from the dynamics observed in external markets. For the polynomial, some relatively more modern methods may also be considered, like, Gaussian processes\cite{rasmussen2003gaussian}, which can be fitted to the data as follows: 
    \begin{equation}
        y(x) \sim \mathcal{N} (m(x; \psi_m), \mathbf{K}(x,x; \psi_C)),
    \end{equation}
    
    \noindent
    where $m$ is the mean function, which has hyperparameters $\psi_m$ that encodes domain knowledge concerning the deterministic component of the dataset, and the covariance matrix $\mathbf{K}$ has hyperparameters $\psi_C$. The mean and covariance functions are to be chosen with domain knowledge of the dataset. The hyperparameters of the proposed model must be marginalised, which refers to integrating out uncertainty\cite{roberts2013gaussian}\footnote{For a given function $p(y,\theta) = p(y|\theta)p(\theta)$, we can obtain the variable of interest $p(y)$ by marginalising over the unknown parameter $\theta$, s.t. $p(y) = \int p(y|\theta)p(\theta)d\theta$.}. Several sophisticated techniques exist to solve the integral approximation, which fits a Gaussian around a maximum-likelihood peak. We leave the specific methodology of curve fitting to be selected by oracle service providers.
    
\subsubsection{Transactional Accounting}
    As described in the preceding subsection, the protocol uses price oracles to map discrete limit order book data to a continuous setting by determining the coefficients of the relevant polynomial. This polynomial was visualised in a price-volume space in the preceding section, but in this section, we switch the axis to a volume-price space, for mathematical convenience.
    
    \begin{definition}[Slot]
        A slot ($s_{\nu}$) is the frequency at which information about the polynomial curve fitted to the external market data is updated in the system, where $s$ is the slot number and $\nu$ is the state of the slot. And each slot is characterised by the start and end time $(t_{s_{0}},t_{s_{T}})$. Naturally, the end of one slot is the beginning of the other, i.e. $t_{{s-1}_{T}}=t_{s_{0}}$.\\
        \\
        The volume traded in each slot is $V^{Y+/-}_{{s_{\nu}}}$, where the $\text{+}$ and $\text{-}$ signify whether indicated volume had been sold or bought, which are recorded disjointly.
    \end{definition}
    
    \noindent
    Using the above definition and new notations, we can state that:
    \begin{equation}
        V^{Y+}_{s_{\nu}} = \sum_{n=t_{s_0}}^{t_{s_\nu}} \Delta \mathcal{I}^\text{Y+}_n,
    \end{equation}
    \noindent
    and,
    \begin{equation}
        V^{Y-}_{s_{\nu}}  =\sum_{n=t_{s_0}}^{t_{s_\nu}} \Delta \mathcal{I}^{Y-}_n,
    \end{equation}
    where $\Delta \mathcal{I}^\text{Y+}_t$ and $\Delta \mathcal{I}^\text{Y-}_t$ are inventory changes in asset $Y$ pool.\\
    \\
    At slot-state, $\nu = 1$, if the externally aggregated ELDF is applied as a pricing rule, the exchange of asset $X$ to asset $Y$ by a trader will lead to the following change in the inventory levels of the market:
    
    \begin{equation}\begin{split}
        \underbrace{(\mathcal{I}^Y_{1}, \mathcal{I}^X_{1})}_\text{before}\rightarrow 
        \overbrace{(\mathcal{I}^Y_{1} + \Delta \mathcal{I}^Y_{2}, \mathcal{I}^X_{1} - (\int_{V^{Y-}_{1_{1}}}^{V^{Y-}_{1_{2}}} (c_2V^{Y^2}+c_1V^Y+c_0)\,dV^Y) )}^\text{after},
    \end{split}\end{equation}
    
    \noindent
    where $V^{Y-}_{1_{2}} = V^{Y-}_{1_{1}} + \Delta \mathcal{I}^Y_2$, $\mathcal{I}^Y_{1}$ represents the inventory level of asset $Y$ before the trade, and $\Delta \mathcal{I}^Y_2$ the change in inventory level of asset $Y$ in the next time step.\\
    \\
    Now, for the calculation of the amount of asset $X$ we can get from the market, we calculate the area under the ELDF (assuming a second-order polynomial) as follows:
    \begin{equation}
        \int_{V^{Y-}_{1_{1}}}^{V^{Y-}_{1_{2}}}( c_2 V^{Y^2}+c_1V^Y+c_0)\,dV^Y = \frac{c_2 \left(-V_{1_{1}}^{{Y-}^{3}}+V_{1_{2}}^{Y-^{3}}\right)}{3}+\frac{c_1 \left(-V_{1_{1}}^{Y-^{2}}+V_{1_{2}}^{{Y-}^{2}}\right)}{2}+  c_0\left(V_{1_{2}}^{Y-} -V_{1_{1}}^{Y-} \right).
    \end{equation}
    
    \noindent
    Solving this equation enables us to ascertain the number of units of asset $Y$, which need to be exchanged if we want to get $M$-many units of the other asset $X$.
    \begin{equation}
        \begin{split}
            \frac{c_2 (-V^{Y-^{3}}_{1_{1}}+V^{Y-^{3}}_{1_{2}})}{3}+ \frac{c_1 (-V^{{Y-}^2}_{1_{1}}+V_{1_{2}}^{{Y-}^2})}{2}+  c_0(V^{Y-}_{1_{2}} -V^{Y-}_{1_{1}})= M.
        \end{split}
    \end{equation}
    
    \noindent
    We solve the equation below for $V_{1_{2}}^{   Y-}$ to obtain feasible roots:
    \begin{equation}
    \begin{split}
        \frac{c_2}{3} V_{1_{2}}^{Y-^3} + \frac{c_1}{2} V_{1_{2}}^{Y-^2} + c_0V_{1_{2},Y-} + \{\frac{-c_2V_{1_{1}}^{Y-^3}}{3} - c_1V_{1_{1}}^{{Y-}^2} - c_0V_{1_1}^{Y-} - M\} = 0.
    \end{split}
    \end{equation}
    
\subsection{Liquidity Provision \& Protection}\label{liquidityprov}
    In this section, we introduce a derivatives-based scheme and an algorithmic accounting asset to facilitate inventory risk transfer, aggregate and effectively route liquidity and value across different asset pools, and enable pLPs to provide single-sided liquidity. Whilst the objective of making liquidity as efficient as possible is intuitive, it is important to also make the best efforts to protect the interest of pLPs, who may not possess the technical expertise and infrastructure required to manage their over (or under) exposure to adverse (or desirable) market changes. Hence, the reason why we introduce a new agent in the system, called secondary LP (sLP), aimed at facilitating the creation of a market for pricing of the excess risk pLPs wish to delegate to more sophisticated market participants. This is one of the novel aspects of DFMM, as it enables users to hedge their unwanted inventory risks by applying the native toolkit, provided by the protocol itself.

    \begin{definition}[Liquidity Network]
        An AMM liquidity network can be defined as a directed graph, $G(V,E)$, where each vertex ($V$) represents paired inventory pools of digital asset submitted to the pool, and each edge represents the liquidity transfers between two paired pools. Each edge is a tuple $(s,r,b)$ where $s$ and $r$ are the paired inventory pools, sending and receiving trades, and $b$ is the value transferred between paired pools. Each path is a sequence of edges, such that the tail of each edge is the head of the next, each vertex is reachable from another if there is a path from the latter to the former, and edges are realised by AMM dynamics. Note, that during an asset exchange in an AMM, an edge can only transfer assets it already owns, i.e. the path capacity during liquidity transfers is limited to the available asset inventory in the respective pools.
    \end{definition}
    
    \begin{figure}
        \begin{center}
        \includegraphics[width=5in]{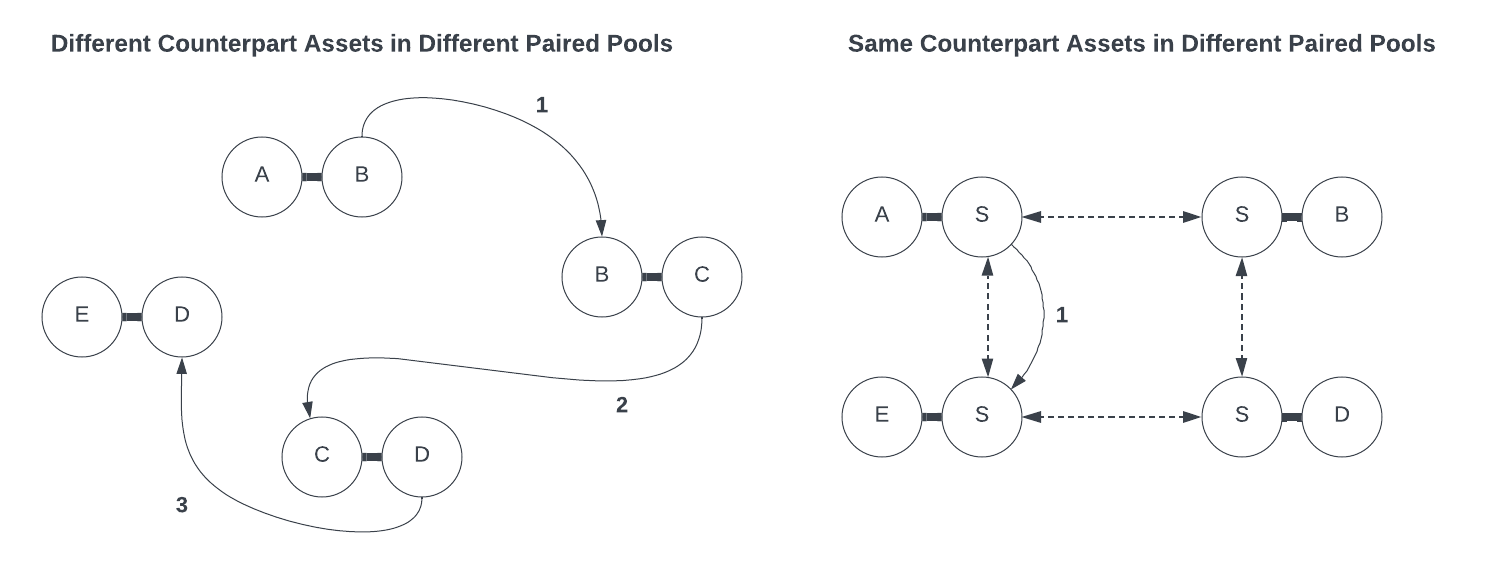}
        \caption{Counterpart Asset.}
        \label{fig:count_asst}
        \end{center}
    \end{figure}

    \noindent
    On the left-hand side in Fig.\ref{fig:count_asst} above, we observe that the paired pools do not have a common counterpart asset (such as $E$ and $A$), necessitating the identification of an optimal path to complete a trade (or swap transaction)\footnote{This design assumes segregation of liquidity pools}. This optimal path should be, for e.g., the one traversing vertices carrying sufficiently high liquidity, as it would help minimise slippage. However, searching for this path for each execution would be computationally inefficient, and thus, we aim to optimise liquidity networks by introducing an intermediary synthetic counterpart asset, which can be thought of as an accounting coin to minimise the length of the routing path.\\
    \\
    A desirable side-effect of this feature is that it would enable pLPs to submit their choice of liquidity to the system, without being compelled to submit the second asset to form a pair, thereby suppressing the risk-adjusted return potential of these market participants, which is a requirement of several popular protocols (e.g. see \cite{adams2021uniswap}).\\
    \\
    Notably, to enhance the stability of the protocol, we seek to denominate the value of the counterpart asset in terms of a suitable\footnote{We leave the definition of a suitable asset (incl. a basket of assets), which can be qualitatively stated to be relatively less volatile and widely adopted, e.g. US Dollar.} asset.

\subsubsection{Internal Accounting}\label{openinventory}
    The protocol's most fundamental objective is to ensure that pLPs can withdraw assets they had deposited (liability for the protocol), or the equivalent value of other assets, which necessitates the following inequality to be continually satisfied in the case of two assets.
    
    \begin{equation}\label{obligation}
         \mathcal{L}^E_{{bid}_t}(\mathcal{I}^X_t) + \mathcal{L}^E_{{bid}_t}(\mathcal{I}^Y_t)  \geq \mathcal{L}^E_{{bid}_t}(\mathcal{I}^X_{{LP}_t}) + \mathcal{L}^E_{{bid}_t}(\mathcal{I}^Y_{{LP}_t}).
    \end{equation}

    \noindent
    This inequality is continually tested as fluctuations in prices lead to changes in the value of each of the aforementioned assets. This can be better understood by considering the schematic diagram in Fig.\ref{fig:bs}, which exemplifies how the protocol's balance sheet evolves after a trade:
    
    \begin{figure}[H]
      \begin{center}
          \includegraphics[width=2.5in]{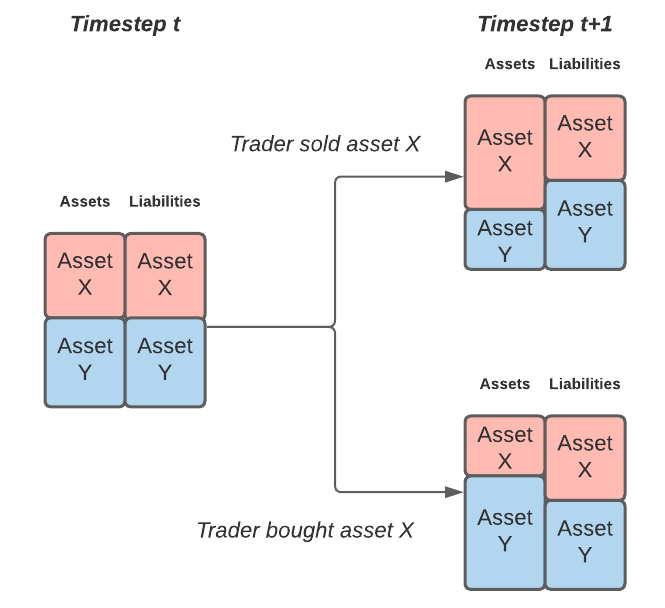}
          \caption{Evolution of the balance sheet over time.}
          \label{fig:bs}
      \end{center}
    \end{figure}

    \noindent
    The mismatch expressed on the right-hand side of the schematic in Fig.\ref{fig:bs} is the market risk the DFMM protocol is exposed to, which is used to define an important measure for the protocol below.
    
    \begin{definition}[Open Inventory]
        The open inventory of an asset $X$, represented by $\tilde{\mathcal{I}_t^X}$, is defined as the difference between the current inventory at time $t$ and the assets owned by LPs, which have been deposited into pools up to that point.
        \begin{equation}
            \tilde{\mathcal{I}_t^X} =  \mathcal{I}^X_t - \mathcal{I}^X_{LP_{t}}.
        \end{equation}
    \end{definition}

    \begin{figure}[H]
      \begin{center}
      \includegraphics[width=4in]{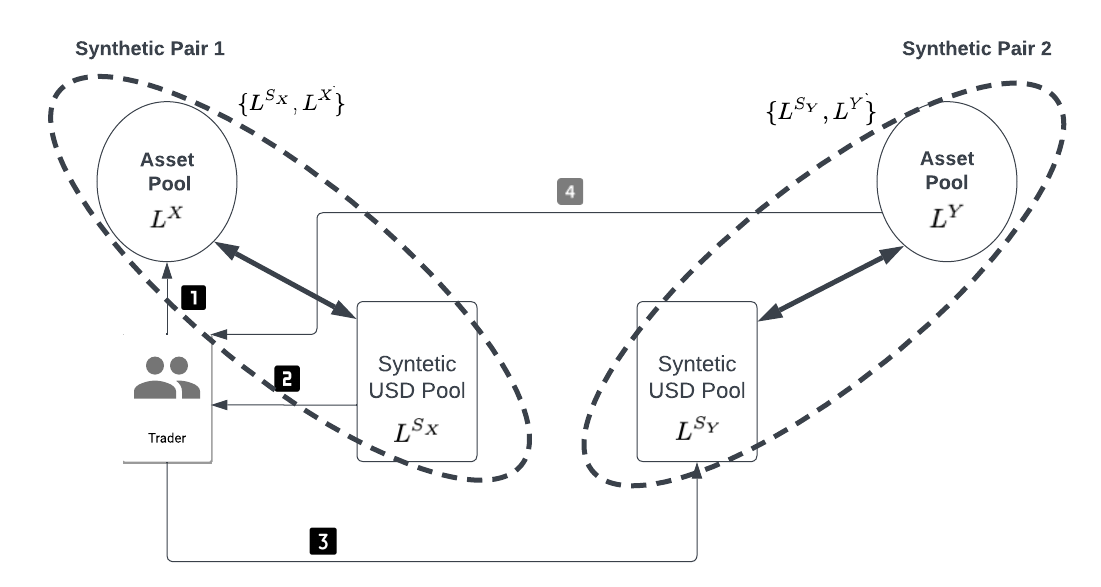}
      \caption{Synthetic Pools.}
      \label{fig:schematicdfmm}
      \end{center}
    \end{figure}

    \noindent
    We now introduce an algorithmic accounting asset (\$S), which we seek to link to the US Dollar and denote the pools comprised of this accounting asset paired with inventory pools $L_X$ and $L_Y$ as $L_{S_X}$ and $L_{S_Y}$, respectively. These are essentially synthetic pairs of pools, i.e., \{$L_X, L_{S_X}$\} and \{$L_Y, L_{S_Y}$\}. The primary purpose of \$S is to introduce a common denomination for the internal reconciliation of trades and measure the nominal value of the open inventory position. To better elucidate the trade flow, consider the situation where an X/Y trade occurs, where the user seeks to sell $X$ to buy $Y$:

    \begin{figure}[H]
        \begin{center}
            \includegraphics[width=4.8in]{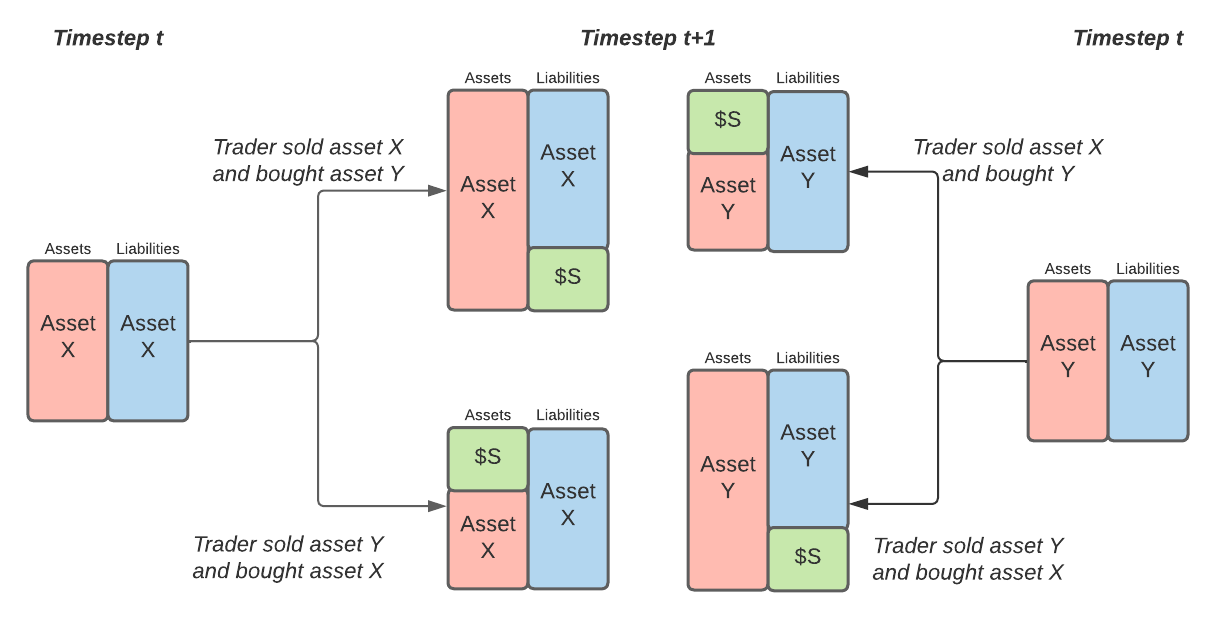}
            \caption{Change of Balance Sheet composition.}
            \label{fig:bs_stable}
        \end{center}
    \end{figure}

    \noindent
    In Fig.\ref{fig:schematicdfmm}, the user first deposits a fixed unit of asset $X$ in the asset pool $L_X$, to which a corresponding number of \$S are synthetically received\footnote{The trader does not receive this asset, but rather it is conducted as an accounting operation.}, from the paired pool $L_{S_X}$. This completes the deposit side of the transaction. Now, to obtain another asset $Y$, the system first moves the equivalent number of \$S received from the pool $L_{S_X}$ to the pool $L_{S_Y}$, which is then converted to the desired asset based on the ongoing pricing rule, of $L_Y$\footnote{We demonstrate the composition of the protocol's balance sheet after the introduction of the accounting asset in Fig.\ref{fig:bs_stable}.}. The volume of the synthetic asset traded in each synthetically paired pool is represented by $T^X_t \in \mathbbm{R}$ and $T^Y_t \in \mathbbm{R}$, which is defined as follows:
    \begin{equation}
        T_t^X= \sum_{i=0}^{t}V_\text{i}^{S_X+} - \sum_{i=0}^{t}V_\text{i}^{S_X-}.
    \end{equation}
    
    \noindent
   where $V_\text{t}^{S_X+}$ and $V_\text{t}^{S_X-}$ are the volume of synthetic asset sold and bought using $\{L_X, L_{S_X}\}$ paired pool upto the timestep $t$.\\
    \\
    Note, that $T^X_t$ also represents the nominal value of the open inventory position of $L^X$ asset pool, $\tilde{\mathcal{I}_t^X}$. If $\tilde{\mathcal{I}_t^X} \neq 0$, we might have a situation where the protocol's $assets < liabilities$, which we must seek to avoid at all costs. We do this by hedging (under constraints) the nominal value of the open inventory positions. The success of this objective ensures that the synthetically created asset can be swapped back to the deposited digital asset, thereby enabling us to view the synthetic asset as a fixed-value algorithmic accounting asset, which we've defined before.

    \begin{definition} [Algorithmic Accounting Asset]
        The algorithmic accounting asset serves as the primary unit of measure (e.g. in USD) for internal accounting operations within the system. These operations enable the system to track open inventory positions and pool holdings, in nominal terms. This accounting asset provides a common denomination for standardising rebalancing processes and reward distribution across various pools that hold different assets, thereby, streamlining operations.
    \end{definition}

    \noindent
    The incorporation of an accounting asset and related risk management program ensures the system’s capability to return liquidity providers (LPs) their initial invested assets, fulfilling the core objective of preserving user inventory. 
    
    The constraints we follow for the hedging program are:
    \begin{equation}
        \begin{cases}
            \forall \mathcal{I}^X_t \geq \mathcal{I}^X_{{LP}_t}, \quad -\mathcal{L}^{X^E}_{{bid}_t}(|\mathcal{I}^X_{{LP}_t} - \mathcal{I}^X_t|) = T^X_t,
             \\ 
            \forall \mathcal{I}^X_t < \mathcal{I}^X_{{LP}_t}, \quad \mathcal{L}^{X^E}_{{ask}_t}(|\mathcal{I}^X_{{LP}_t} - \mathcal{I}^X_t |) = T^X_t.
 
        \end{cases}
    \end{equation}

    \noindent
    In the forthcoming sections, we introduce a new agent to the system and present a non-linear digital finance instrument to enable us to ensure that the aforementioned objectives are assured.
    
\subsubsection{Secondary Liquidity Providers (sLP)}
    In this subsection, we introduce a new agent, who is technically and technologically, a more skilled market participant, and seeks to participate in complex and nuanced risks, which pLPs may wish to offset.

    \begin{definition}[Secondary Liquidity Provider (sLP)]
        Secondary liquidity providers (sLP) are market participants similar to pLPs, as defined above. However, unlike pLPs, sLPs do not deposit digital assets into liquidity pools. Instead, they deposit assets into a specialized margin vault as collateral, offering limited protection to passive pLPs. This protection is extended using non-linear financial instruments (derivatives) and involves making a market by posting bid and ask prices in the protocol\footnote{The protocol aims to incentivize the submission of collateral to both the long and short sides of the vaults to promote a meaningful market.}.
    \end{definition}
    
\begin{figure}[H]
        \begin{center}
            \includegraphics[width=3.8in]{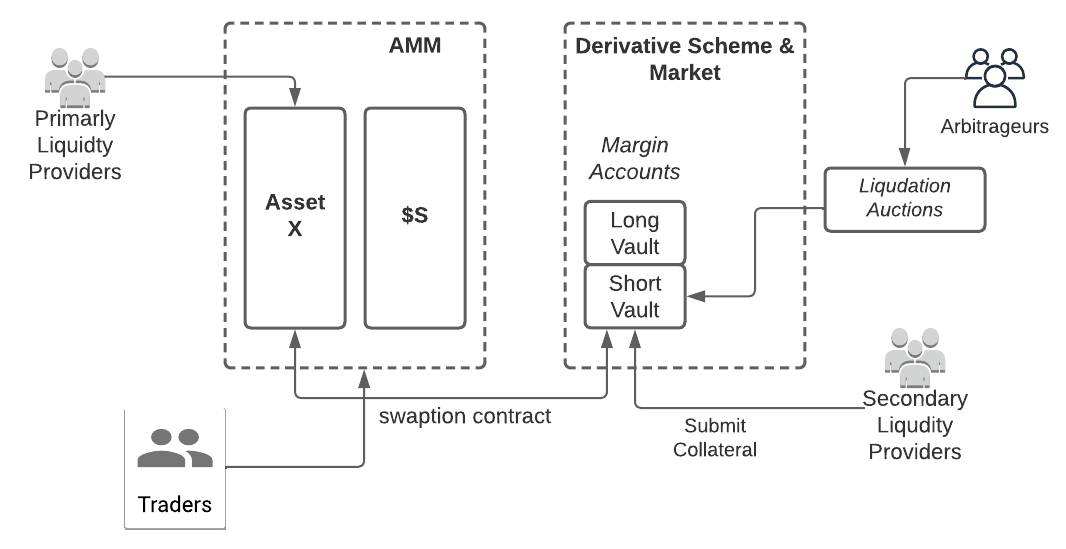}
            \caption{Schematics of long and short sLP vaults.}
            \label{fig: sLP vaults}
        \end{center}
    \end{figure}

    \noindent
    We denote the total collateral available in long and short vaults\footnote{Long (and short) vaults of an asset are margin accounts where sLPs can lock-in digital assets, to express a firm interest offering (bidding) a non-linear financial instrument aimed at enabling pLPs risk to be offset their market risk, through the protocol, on either long or short side.} associated with the pool of asset $X$ at time step $t$ by $\mathcal{C}_{{long}_t}^X$ and $\mathcal{C}_{{short}_t}^X$, respectively. Once deposited, this collateral is used by the protocol to execute a derivative trade with the sLP, to transfer the transitory risk of currency mismatches. This mechanism is explained in the schematics presented in Fig.\ref{fig: sLP vaults}, which follows.

    \begin{definition}[Collateralisation Rate]
        The collateralisation rate for a long or short position in asset X, denoted as $\varrho_{long}^X$ or $\varrho_{short}^X$ respectively, represents the percentage of unhedged open inventory that must be held in the vault as collateral to mitigate the inventory risk associated with the position.
    \end{definition}

    \noindent
    The aforementioned measure enables us to calculate the maximum amount of open inventory of asset $X$, which can be supported by collateral available in margin vaults, as: $\frac{\mathcal{C}^X_{{short}_t}}{\varrho_{short}^X}$ and $\frac{\mathcal{C}^X_{{long}_t}}{\varrho_{long}^X}$.    
    
    \begin{definition}[Bundle]
        A bundle is a tuple, comprised of spot and derivative position.
    \end{definition}

    \begin{definition}[Complete Bundle]
        The protocol is said to have a complete bundle for a particular asset pool, if the net point-in-time delta of the spot and derivative is zero, i.e. it has no directional exposure to movement in the spot price - enabling the protocol to honour its obligations (see equation \ref{obligation}) to pLPs.
        \begin{equation}
            \Delta_{t}^{\text{Bundle}_X} = \Delta^{pLP}_{X_t} + \Delta^{sLP}_{X_t}.
        \end{equation}
    \end{definition}

\subsubsection{Digital Swaption}
    In the preceding sections, we have introduced the essential components of the DFMM ecosystem, including key agents, the algorithmic accounting asset, and inventory management processes. In this section, we explore how a non-linear financial instrument is used for DFMM's risk management framework, which offers asset owners a means for desired risk mitigation, while also providing sophisticated investors with opportunities to engage in derivative markets by constructing intricate risk-reward portfolios.
    
    \begin{definition}[Digital Swaption]
        A digital swaption is a non-recourse, non-linear, financial instrument, bestowing its holder the right, but not enforcing an obligation, to enter into a total return swap, starting at a pre-specified date in the future, for a fixed maturity and rate, set at the time of inception. 
    \end{definition}
    
\begin{figure}[H]
      \begin{center}
          \includegraphics[width=3.9in]{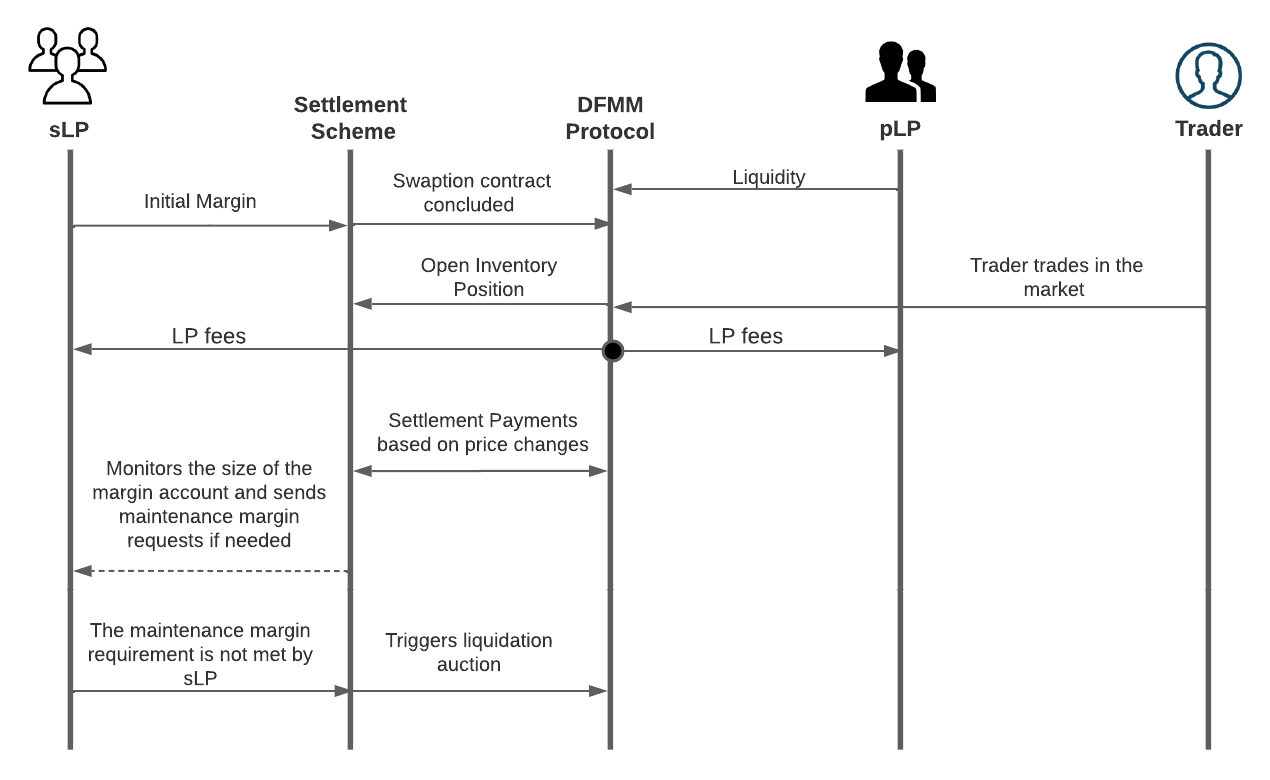}
          \caption{sLP protection mechanism.}
          \label{fig:swaption_scheme}
      \end{center}
    \end{figure}

    \noindent
    In the definition of the digital swaption above, the party (sLP) exposed agrees to absorb undesirable exposure of the pLP, up to a preset threshold of $r^u$. In return, the DFMM pays a premium $\psi(\mathbf{X}, r^u)$, which is affected by the entire returns distribution of the asset $X$ and the upper threshold set based on tolerance to losses stemming from adverse moves. In many cases, $r^u = 0$ since holders seek complete assurance against adverse moves. We refer to it as a `digital' swaption because it is based on digital assets, as opposed to interest-bearing instruments like bonds, or dividend-yielding stock, and its settlement is hard-coded in the blockchain. Furthermore, since the introduced non-linear financial instrument is non-recourse, it actually behaves like a put-spread to manage long exposure to the digital asset underlying the swap and a call-spread to manage short exposure to the digital asset underlying the swap.\\
    \\
    In the schematic presented in Fig.\ref{fig:swaption_scheme}, we describe the interaction between different agents of the system and DFMM's role in intermediating such transactions. 
    
    \noindent
    Simply put, if as the result of trades in an epoch, the token inventory decreases from the level provided by pLPs (i.e., pLPs are exposed to open inventory positions), then the protocol is taking a long position where the sLPs who deposit collateral in the short vault, are taking the variable leg, and the protocol is covering the fixed leg. Similarly, if the opposite is true and the token inventory has increased as a result of the trading activity in an epoch, then protocol takes a short position covering the variable leg and receiving the fixed leg. This new desired exposure is obtained by the protocol entering into the above-described derivative contract with an sLP, who has deposited collateral in the relevant long (or short) vault. In summary:
    
    \begin{equation}
        \left\{
            \begin{array}{lr}
                \forall \mathcal{I}^X_t - \mathcal{I}^X_{LP_t} > 0, \text{DFMM is variable leg payer} \\
                \forall \mathcal{I}^X_t - \mathcal{I}^X_{LP_t} < 0, \text{DFMM is fixed leg payer} \\
                \forall \mathcal{I}^X_t - \mathcal{I}^X_{LP_t} = 0, \text{No swaption position has been exercised}. 
            \end{array}
        \right\}
    \end{equation}
    
    \noindent
    To hedge this risk, the system enters into a digital contract with sLP, where the notional of the contract is based on the current open inventory position in the system ($|\mathcal{I}^X_t - \mathcal{I}^X_{LP_t}|$). The fixed cashflow stream in the newly initiated contract is based on the liquidity density function of the previous epoch $\mathcal{L}_{t-1}^{E_X}(|\mathcal{I}^X_t - \mathcal{I}^X_{LP_t}|)$. On the other hand, the variable cashflow stream is based on the liquidity density function of the current epoch  $\mathcal{L}_{t}^{E_X}(|\mathcal{I}^X_t - \mathcal{I}^X_{LP_t}|)$. If $\mathcal{L}_{t}^{E_X}(|\mathcal{I}^X_t - \mathcal{I}^X_{LP_t}|) > \mathcal{L}_{t-1}^{E_X}(|\mathcal{I}^X_t - \mathcal{I}^X_{LP_t}|)$, where the settlement amount ($SA_t$) is paid by variable leg payer to the fixed leg owner, and can be calculated as follows: 
    \begin{equation}
        SA_t = |\mathcal{I}^X_t - \mathcal{I}^X_{LP_t}| \{\frac{\mathcal{L}_{t}^{E_X}(|\mathcal{I}^X_t - \mathcal{I}^X_{LP_t}|)}{\mathcal{L}_{t-1}^{E_X}(|\mathcal{I}^X_t - \mathcal{I}^X_{LP_t}|)} -1\}.
    \end{equation}
    
    \noindent
    \underline{\textbf{Residual Risk}}: The derivative position and its users are naturally exposed to counterparty risk, but these are mitigated in a fully automated digital finance platform. The residual risk that remains is one that arises from the non-recourse nature of sLP vaults, which means that the coverage is tantamount to limited protection through a put or call spread and not a vanilla put or call option. To help minimise the gap, DFMM deploys the following levers:
    
    \begin{itemize}
        \item Settlement: Optimally shorter settlement frequency is used to maximally minimise the difference between paper and realised profits (or losses) stemming from a derivative position.
        \item Margin: sLPs are required to continuously satisfy margin requirement ($\mathcal{C}^X_t > \epsilon$) for an asset, where $\epsilon$ is the floor below which positions are liquidated, and a decentralised governance framework can be used to set requirements specific to idiosyncratic assets.
    \end{itemize}

\subsection{Local Price Formation}\label{localprice}
    Given the presence of two distinct, but interconnected venues, reflecting the price of the same asset - i.e. the local price within DFMM and aggregated prices fetched from external markets, and the assumption that one of the ``market is always right", means that there has to be a mechanism enabling synchronisation of the local market with the external market. Therefore in this section, we discuss how information from the external market is used to augment internally available data, enabling the protocol to adapt its local liquidity density function to incentivise rebalancing flows.

    \begin{definition}[Rebalancing Premium]
        Rebalancing premium is the price difference of an asset, between the local and external markets, and can be seen as the theoretical upper bound to the arbitrageable value traders can seek to monetise, to help rebalance inventory.
    \end{definition}

    \noindent
    Note, that in the state where the DFMM inventory level is equal to the inventory level submitted by LP i.e. $\mathcal{I}^X_{{LP}_t}=\mathcal{I}^X_t$ there should be no arbitrageable value in the system, $\mathcal{L}_t^{E_X}(V^X_t)=\mathcal{L}_t^{L_X}(V^X_t)$. However, if there is a unsustainable surplus of inventory in the system, i.e. $\mathcal{I}^X_{{LP}_t}<\mathcal{I}^X_t$, the system should incentivise arbitrageurs to buy from the internal market and sell in the external market and vice versa if $\mathcal{I}^X_{{LP}_t}>\mathcal{I}^X_t$, the system should incentivise arbitrageurs to buy from the external market and sell in the internal market.

    \begin{definition}[Rebalancing Premium Function]
        The rebalancing premium function $\mathcal{R}^X_t$, where is the trading volume of the asset $X$, determines the optimal arbitrageable value, which will bring the inventory level of the DFMM back to an optimum level, without overly compromising the interest of traders or the protocol. This function seeks to strike the right balance between interests of traders and the protocol, seeking to continually avoid a situation where pLPs have large unhedged poisitions, leading to depletion of reserves, halting of trades, or detrimental to the protocol's risk objectives.
    \end{definition}
    
    \begin{figure}
        \begin{center}
            \includegraphics[width=3.5in]{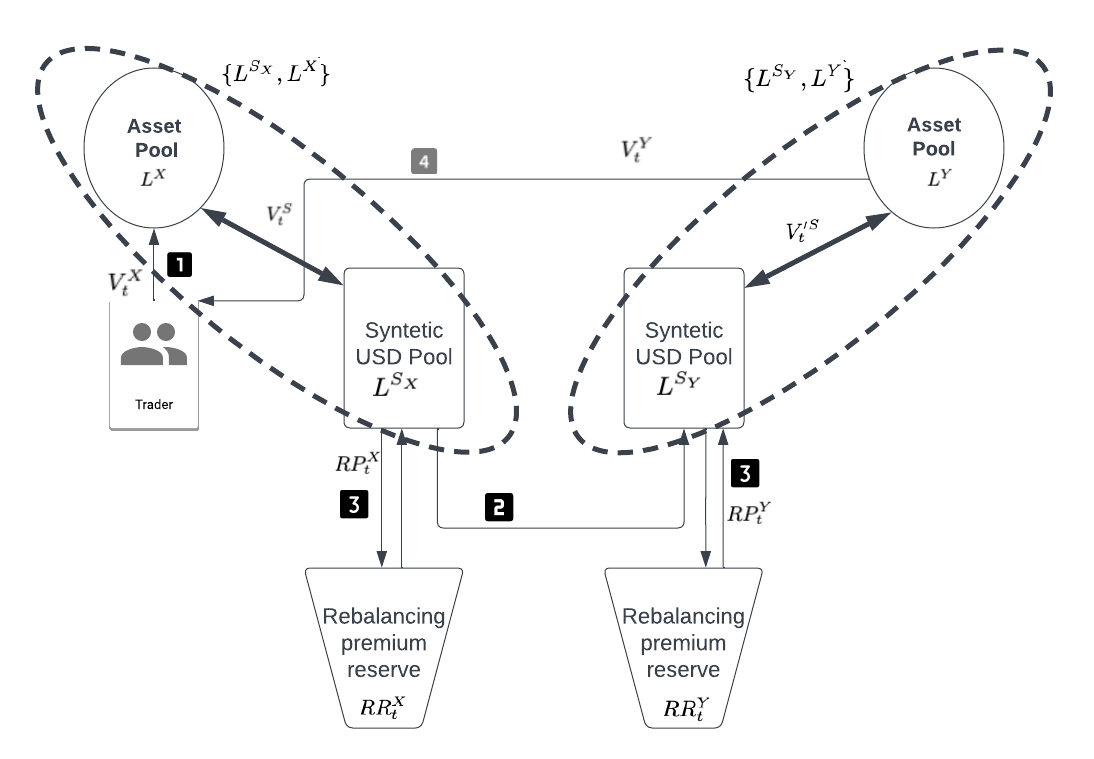}
            \caption{Schematics of local market price formation}
      \end{center}
    \end{figure}
    
    \noindent
    Consider that a trader aims to exchange $V_t^X$ units of asset $X$ to asset $Y$, and wishes to ascertain the number of units of asset $Y$, that trader will get through the exchange in DFMM, i.e. calculate $V_t^Y$. Upon submission of $V_t^X$ units of the asset $X$ to the pool, it is converted to USD using the ELDF\footnote{The derived ELDF would show the aggregated (external) volume-price curve of the asset $X$ in terms of accounting asset(\$S).}.
    
    \begin{equation}
        \mathcal{L}^{E_X}_{{bid}_t}(V^X_t)= V^S_t
    \end{equation}
    
    \noindent
    From an internal accounting perspective, if there was no re-balancing premium, then $V^S_t$ amount of USD would be withdrawn from \$S pool paired to asset $X$, $L^{S_X}$. Similarly, if there was no re-balancing premium, $V_t^S$ amount of USD acquired would be submitted to the \$S pool paired with asset Y's pool ($L^{S_Y}$) to withdraw asset $Y$. The amount withdrawn from the pool $L^{S_X}$ and submitted to the pool $L^{S_Y}$ would be different (from $V_t^{S}$) if there is a rebalancing premium.    
    To trace the rebalancing needs, we can apply $T^X_t$, which, as discussed in Sec. \ref{openinventory} reflects the open inventory position in nominal terms.\\
    \\
    Now, based on the value of $T_t^\text{X}$, the system adopts the following objectives. If,
    \begin{itemize}
        \item $T_t^{X}>0$: The system incentivises the purchase of \$S from the accounting asset pool ($L^{S_X}$). As $L^{S_X}$ is paired with $L^X$ (asset $X$ pool), this objective can be equivalently understood as the incentivisation of sales of asset $X$ to $L^{X}$ pool.
        \item $T_t^{X}<0$: The system should incentivises sale of \$S to the synthetic pool $L^{S_X}$, or equivalently incentivises the acquisition of $X$ asset from $L^X$ pool.
        \item $T_t^{X}=0$: The system is in an optimal state.
    \end{itemize}

    \noindent
    Furthermore, we seek the rebalancing premia function to have the following mathematical properties:
    \begin{axiom}
        \label{ax:Third} The function should be an increasing convex function.
    \end{axiom}
    
    \begin{axiom}
        \label{ax:Fourth} The function should satisfy the boundary condition, that $\lim_{\mathcal{I}^X_t - \mathcal{I}_{{LP}_t}^X \to \infty} f(x) \rightarrow \infty$.
    \end{axiom}

    \noindent
    Whilst the first axiom ensures that the arbitrageable value increases at an incrementally higher rate with the imbalance, the second axiom becomes straightforward for an increasing function.\\
    \\
    Note, that the rebalancing premia function($\mathcal{R^\text{X}}_t$), which maps $T_t^\text{X}$ to the notional amount (in USD) of total rebalancing premium available for a pair (such as $\{L^{S_X},L^X\}$), is used to incentivise or penalise trades in a certain direction.\\
    \\
    We now defined our rebalancing function to be of the form:
    \begin{equation}
        \begin{cases}
           \mathcal{R^\text{X}}_{{RHS}_t}(T_t^X) = T_t^X(T_t^X+\mathcal{A}^X_{{RHS}_t}) \times \mathcal{D}^X_{{RHS}_t}  \forall T_t^X \geq 0 \\       \mathcal{R^\text{X}}_{{LHS}_t}(T_t^X) = -T_t^X(-T_t^X+\mathcal{A}^X_{{LHS}_t}) \times \mathcal{D}^X_{{LHS}_t}  \forall  T_t^X < 0
        \end{cases}
    \end{equation}

    \noindent
    where, we introduce subscripts $RHS$ and $LHS$, to refer to the rebalancing function when $T_t^X \geq 0$ and $T^X_t < 0$, $\mathcal{A}_{RHS}$ and $\mathcal{A}_{LHS}$ are aggressiveness parameters of the system, determined using the auction process, and $\mathcal{D}$ is premium discount received once a derivative-based protection mechanism is paired.\\
    \\
    We can now calculate the nominal value of the rebalancing premium ($RP_t^{X}$), as follows:
    
    \begin{equation}
    RP_t^{X} =
        \begin{cases}
            \mathcal{R}^\text{X}_{{RHS}_t}(T_{t-1}^{X}) -\mathcal{R}^{X}_{{RHS}_t}(T_t^{X})  &\forall T_{t-1}^X \geq 0, T_t^X \geq 0 \\       \mathcal{R}^X_{{RHS}_t}(T^X_{t-1})-\mathcal{R}^\text{X}_{{LHS}_t}(T_t^X) &\forall T_{t-1}^X \geq 0, T_t^X < 0\\
            \mathcal{R}^X_{{LHS}_t}(T^X_{t-1})-\mathcal{R}^\text{X}_{{RHS}_t}(T_t^X) &\forall T_{t-1}^X < 0, T_t^X > 0\\
            \mathcal{R^{X}}_{{LHS}_t}(T_{t-1}^{X}) - \mathcal{R}^\text{X}_{{LHS}_t}(T_{t}^{X}) & \forall T_{t-1}^X < 0, T_t^X \leq 0.\\    
        \end{cases}
    \end{equation}

    \noindent
    Similarly, $RP^Y_t$ could be calculated same way, where, $T_t^{X}=T_{t-1}^{X} - V^{'S}_t$, and $T_t^{Y}=T_{t-1}^{Y} + V^{'S}_t$  and $V'^S_t$ represents change of the size of accounting asset pools.\\
    \\    
    Note, that since $\mathcal{R}_{t}^{X}$ is an increasing function, any trade which increases $|T_t^\text{X}|$ compared with the previous system-state leads to a positive rebalancing premium ($RP^\text{X}_t>0$), which is submitted to the rebalancing premium reserve($RR^{X}_t + RP^\text{X}_t$) at the expense of the trader whose trade lead to the change. And similarly, any trade which decreases $|T_t^{X}|$ relative to the previous epoch leads to a negative rebalancing premium ($RP^\text{X}_t<0$), that is paid from rebalancing premium reserve($RR^\text{X}_t + RP^\text{X}_t$) to benefit the trader.\\
    \\
    In the equation below, we state the condition that must be satisfied whilst incorporating the rebalancing premia to the nominal amount($V^{S}_t$):
    \begin{equation}
        V^S_t = V_t^{'S} + RP_t^{X} + \theta_t \times V^S_t,
    \end{equation}
    
    \noindent
    where $V_t^{'S}$ is the nominal \$S after adjustment, and $\theta_t$ is the AMM fee.\\
    \\
    Recall, that every trade includes two different synthetic trades (and two different digital assets $X$ and $Y$), assuming the change in two different synthetic USD pools ($L^{S_X}$ and $L^{S_Y}$) and correspondingly in 2 different rebalancing premium reserves($RR^{X}$ and $RR^Y$). If both of the pools are adjusted by the same amount of the algorithmic accounting asset $V_t^{'S}$, the following should be true:
    
    \begin{equation}
         V^S_t = V_t^{'S} + RP_t^X+ RP_t^Y +\theta_t \times V^S_t,
    \end{equation}
    
    \noindent
    where, $\theta_t$ is the AMM fees charged from the traders by the market maker for the service.\\
    \\
    The only unknown in the equation above is $V_t^{'S}$, solving the equation for $V_t^{'S}$ will help us to derive the new state of $T_t^X$ and  $T_t^Y$ as well as the rebalancing premium generated in two different pools, $RP_t^X$ and $RP_t^Y$. We can apply conditional logic to find a solution, where we first test to see in which interval $T_t^X$ and $T_t^Y$ will fit, and then based on tested logic we apply the corresponding interval to find a solution.\\
    \\
    Now in Algorithm 1, we calculate $V_t^{'S}$ when a trader swaps asset $Y$ for $X$, leading to an increase in $T^X_{t-1}$ and decrease in $T^Y_{t}$.
    
    \begin{algorithm}[H]
     \scriptsize
        \floatname{algorithm}{Algorithm 1}
        \renewcommand{\thealgorithm}{}
        \caption{Rebalancing premium adjusted price.}. 
        \label{protocol1}
        \begin{algorithmic}[1]
%1   

        \State \textbf{if}  
        $T^X_{t-1}  \geq 0$ \text{and}  $T^Y_{t-1}  \leq 0$\; \textbf{then}
        \State \textcolor{white}{......} $V^{'S}_t - \mathcal{R}^{X}_{{RHS}_t}(T_{t-1}^{X}) +\mathcal{R^{X}}_{{RHS}_t}(T_t^{X}) -\mathcal{R^{Y}}_{{LHS}_{t}}(T_{t-1}^{Y}) + \mathcal{R}^\text{Y}_{{LHS}_{t}}(T_{t}^{Y}) + \mathcal{F}_t \times V^S_t = V^S_t$
        \State \textcolor{white}{......} \textbf{solve for} $V^{'S}_t$

%2        
        % \State 
        \State \textbf{elif }$T^X_{t-1}  \geq 0$ \text{and}  $T^Y_{t-1}  > 0$\; \textbf{then}
            \State 
        \State \textcolor{white}{......}  \textbf{if} $T^Y_{t-1} - \mathcal{R}^\text{X}_{{RHS}_t}(T_{t-1}^{X}) +\mathcal{R^{X}}_{{RHS}_t}(T_{t-1}^{X}+T^Y_{t-1}) + \mathcal{F}_t \times V^S_t > V^S_t$ \textbf{then}
         \State \textcolor{white}{...........} $V^{'S}_t - \mathcal{R}^\text{X}_{{RHS}_t}(T_{t-1}^{X}) +\mathcal{R^{X}}_{{RHS}_t}(T_t^{X}) -  \mathcal{R}^\text{Y}_{{RHS}_t}(T_{t-1}^{Y}) +\mathcal{R^{Y}}_{{RHS}_t}(T_t^{Y}) + \mathcal{F}_t \times V^S_t = V^S_t $
    
        \State \textcolor{white}{...........}   \textbf{solve for} $V^{'S}_t$

%3        
            % \State 
         \State \textcolor{white}{......}  \textbf{if} $T^Y_{t-1} - \mathcal{R}^\text{X}_{{RHS}_t}(T_{t-1}^{X}) +\mathcal{R^{X}}_{{RHS}_t}(T_{t-1}^{X}+T^Y_{t-1}) + \mathcal{F}_t \times V^S_t < V^S_t$ \textbf{then}
           \State \textcolor{white}{...........} $V^{'S}_t - \mathcal{R}^\text{X}_{{RHS}_t}(T_{t-1}^{X}) +\mathcal{R^{X}}_{{RHS}_t}(T_t^{X}) - \mathcal{R^{Y}}_{{RHS}_t}(T^Y_{t-1})+ \mathcal{R^{Y}}_{{LHS}_t}(T^Y_t)+ \mathcal{F}_t \times V^S_t = V^S_t$
          \State \textcolor{white}{...........}   \textbf{solve for} $V^{'S}_t$

 %4         
              % \State 
         \State \textcolor{white}{......}  \textbf{else}
           \State \textcolor{white}{...........}  $V^{'S}_t =T^Y_{t-1} $ 
              % \State 

%5              
     \State \textbf{elif }$T^X_{t-1} < 0$ \text{and}  $T^Y_{t-1}  \leq 0$\; \textbf{then}
               % \State 
         \State \textcolor{white}{......}  \textbf{if} $-T^X_{t-1} - \mathcal{R}^\text{Y}_{{LHS}_t}(T_{t-1}^{Y}) +\mathcal{R^{Y}}_{{LHS}_t}(T_{t-1}^{X}+T^Y_{t-1}) + \mathcal{F}_t \times V^S_t > V^S_t$ \textbf{then}
         \State \textcolor{white}{...........}   $V^{'S}_t -  \mathcal{R^{X}}_{{LHS}_t}(T_t^{X}) + \mathcal{R}^\text{X}_{{LHS}_t}(T_{t-1}^{X}) - \mathcal{R^{Y}}_{{LHS}_t}(T_t^{Y}) + \mathcal{R}^\text{Y}_{{LHS}_t}(T_{t-1}^{Y}) + \mathcal{F}_t \times V^S_t = V^S_t$
        \State \textcolor{white}{...........}   \textbf{solve for} $V^{'S}_t$     
        % \State 

%6          
           \State \textcolor{white}{......}  \textbf{elif} $-T^X_{t-1} - \mathcal{R}^\text{Y}_{{LHS}_t}(T_{t-1}^{Y}) +\mathcal{R^{Y}}_{{LHS}_t}(T_{t-1}^{X}+T^Y_{t-1}) + \mathcal{F}_t \times V^S_t < V^S_t$ \textbf{then}   
           \State \textcolor{white}{...........}   $V^{'t}_S + \mathcal{R}^X_{{LHS}_t}(T^X_{t-1}) - \mathcal{R}^X_{{RHS}_t}(T_t^X) - \mathcal{R^{Y}}_{{LHS}_t}(T_t^{Y}) + \mathcal{R}^\text{Y}_{{LHS}_t}(T_{t-1}^{Y}) + \mathcal{F}_t \times V^S_t = V^S_t$    
            \State \textcolor{white}{......}  \textbf{else}    
             \State \textcolor{white}{...........}  $V^{'S}_t =-T^X_{t-1} $  

     \State \textbf{else } 
               % \State 
    \State \textcolor{white}{......}  \textbf{if} $|T^X_{t-1}| \geq |T^Y_{t-1}|$
     \State \textcolor{white}{...........} \textbf{if} $T^Y_{t-1} +\mathcal{R^{X}}_{{LHS}_t}(T_{t-1}^X + T^Y_{t-1}) - \mathcal{R}^\text{X}_{{LHS}_t}(T_{t-1}^{X}) + \mathcal{F}_t \times V^S_t >  V^S_t$ \textbf{then}  
     \State \textcolor{white}{...............}$ V^S_t + \mathcal{R^{X}}_{{LHS}_t}(T_t^{X}) - \mathcal{R}^\text{X}_{{LHS}_t}(T_{t-1}^{X}) - \mathcal{R}^\text{Y}_{{RHS}_t}(T_{t-1}^{Y}) +\mathcal{R^{Y}}_{{RHS}_t}(T_t^{Y})+ \mathcal{F}_t \times V^S_t =  V^S_t$    
     \State \textcolor{white}{...............}    \textbf{solve for} $V^{'S}_t$
     \State \textcolor{white}{...........} \textbf{elif} $T^Y_{t-1} +\mathcal{R^{X}}_{{LHS}_t}(T_{t-1}^X + T^Y_{t-1}) - \mathcal{R}^\text{X}_{{LHS}_t}(T_{t-1}^{X}) + \mathcal{F}_t \times V^S_t <  V^S_t$ \textbf{and} $-T^X_{t-1} - \mathcal{R}^Y_{{RHS}_t}(T_{t-1}^Y) + \mathcal{R}^{Y}_{{LHS}_t}(T_{t-1}^{Y}+T_{t-1}^X)+ \mathcal{F}_t \times V^S_t >  V^S_t$ \textbf{then} 
      \State \textcolor{white}{...............}$ V^{'S}_t + \mathcal{R^{X}}_{{LHS}_t}(T_t^{X}) - \mathcal{R}^\text{X}_{{LHS}_t}(T_{t-1}^{X}) -  \mathcal{R}^\text{Y}_{{RHS}_t} (T_{t-1}^{Y})+\mathcal{R}^\text{Y}_{{LHS}_t}(T_t^{Y})+ \mathcal{F}_t \times V^S_t =  V^S_t$    
     \State \textcolor{white}{...........} \textbf{elif} $T^Y_{t-1} +\mathcal{R^{X}}_{{LHS}_t}(T_{t-1}^X + T^Y_{t-1}) - \mathcal{R}^\text{X}_{{LHS}_t}(T_{t-1}^{X}) + \mathcal{F}_t \times V^S_t <  V^S_t$ \textbf{and} $-T^X_{t-1} - \mathcal{R}^Y_{{RHS}_t}(T_{t-1}^Y) + \mathcal{R}^{Y}_{{LHS}_t}(T_{t-1}^{Y}+T_{t-1}^X)+ \mathcal{F}_t \times V^S_t <  V^S_t$ \textbf{then} 
      \State \textcolor{white}{...............}$ V^{'S}_t -\mathcal{R}^X_{{LHS}_t}(T^X_{t-1}) + \mathcal{R}^X_{{RHS}_t}(T^X_t)- \mathcal{R}^Y_{{RHS}_{t}}(T^Y_{t-1}) + \mathcal{R}^Y_{{LHS}_t}(T^Y_t)  + \mathcal{F}_t \times V^S_t =  V^S_t$  
    \State \textcolor{white}{...............}  \textbf{solve for}   $V^{'S}_t$
    \State \textcolor{white}{...........} \textbf{elif} $T^Y_{t-1} -\mathcal{R^{X}}_{{LHS}_t}(T_{t-1}^X + T^Y_{t-1}) + \mathcal{R}^\text{X}_{{LHS}_t}(T_{t-1}^{X}) + \mathcal{F}_t \times V^S_t =  V^S_t$  \textbf{then}
    \State \textcolor{white}{...............}    $V^{'t}_X = T^Y_{t-1} $
    \State \textcolor{white}{...........} \textbf{elif} $-T^X_{t-1} +\mathcal{R}^Y_{RHS_t}(T^Y_{t-1}) - \mathcal{R}^Y_{LHS_t}(T^Y_{t-1} + T^X_{t-1}) + \mathcal{F}_t \times V^S_t =  V^S_t$ \textbf{then}
    \State \textcolor{white}{...............}    $V^{'t}_X = -T^X_{t-1} $
    \State \textcolor{white}{......}  \textbf{if} $|T^X_{t-1}| < |T^Y_{t-1}|$ 
     \State \textcolor{white}{...........} \textbf{if} $-T^X_{t-1} -\mathcal{R}^\text{Y}_{{RHS}_t}(T_{t-1}^{Y}) +\mathcal{R^{Y}}_{{RHS}_t}(T_{t-1}^{Y}+T^X_{t-1})+ \mathcal{F}_t \times V^S_t >  V^S_t$ \textbf{then}  
    \State \textcolor{white}{...............}$ V^{'S}_t + \mathcal{R^{X}}_{{LHS}_t}(T_t^{X}) - \mathcal{R}^\text{X}_{{LHS}_t}(T_{t-1}^{X})- \mathcal{R}^\text{Y}_{{RHS}_t}(T_{t-1}^{Y}) +\mathcal{R^{Y}}_{{RHS}_t}(T_t^{Y})+ \mathcal{F}_t \times V^S_t =  V^S_t$    
    \State \textcolor{white}{...............}    \textbf{solve for} $V^{'S}_t$   
    \State \textcolor{white}{...........} \textbf{if} $-T^X_{t-1} -\mathcal{R}^\text{Y}_{{RHS}_t}(T_{t-1}^{Y}) +\mathcal{R^{Y}}_{{RHS}_t}(T_{t-1}^{Y}+T^X_{t-1})+ \mathcal{F}_t \times V^S_t <  V^S_t$  \textbf{and} $T^Y_{t-1} - \mathcal{R}^X_{{LHS}_t}(T^X_{t-1}) + \mathcal{R}^X_{{RHS}_t}(T^Y_{t-1}+T^X_{t-1})+ \mathcal{F}_t \times V^S_t >  V^S_t$ \textbf{then}
    \State \textcolor{white}{...............}$ V^{'S}_t -  \mathcal{R}^X_{{LHS}_t}(T^X_{t-1}) +  \mathcal{R}^X_{{RHS}_t}(T^X_t) - \mathcal{R}^Y_{{RHS}_t}(T^Y_{t-1})+\mathcal{R}^Y_{RHS_t}(T^Y_t) + \mathcal{F}_t \times V^S_t =  V^S_t$
    \State \textcolor{white}{...............}    \textbf{solve for} $V^{'S}_t$ 
     \State \textcolor{white}{...........} \textbf{elif} $-T^X_{t-1} -\mathcal{R}^\text{Y}_{{RHS}_t}(T_{t-1}^{Y}) +\mathcal{R^{Y}}_{{RHS}_t}(T_{t-1}^{Y}+T^X_{t-1})+ \mathcal{F}_t \times V^S_t <  V^S_t$  \textbf{and} $T^Y_{t-1} - \mathcal{R}^X_{{LHS}_t}(T^X_{t-1}) + \mathcal{}al{R}^X_{{RHS}_t}(T^Y_{t-1}+T^X_{t-1})+ \mathcal{F}_t \times V^S_t <  V^S_t$ \textbf{then}  
    \State \textcolor{white}{...............}$ V^{'S}_t -\mathcal{R}^X_{{LHS}_t}(T^X_{t-1}) + \mathcal{R}^X_{{RHS}_t}(T^X_t)- \mathcal{R}^Y_{{RHS}_{t}}(T^Y_{t-1}) + \mathcal{R}^Y_{{LHS}_t}(T^Y_t)  + \mathcal{F}_t \times V^S_t =  V^S_t$  
    \State \textcolor{white}{...............}    \textbf{solve for} $V^{'S}_t$ 
    \State \textcolor{white}{...........} \textbf{if} $-T^X_{t-1} -\mathcal{R}^\text{Y}_{{RHS}_t}(T_{t-1}^{Y}) +\mathcal{R^{Y}}_{{RHS}_t}(T_{t-1}^{Y}+T^X_{t-1})+ \mathcal{F}_t \times V^S_t =  V^S_t$
    \State \textcolor{white}{...............}   $V^{'S}_t = -T^X_{t-1} $ 
    \State \textcolor{white}{...........} \textbf{if} $T^Y_{t-1} - \mathcal{R}^X_{{LHS}_t}(T^X_{t-1}) + \mathcal{R}^X_{{RHS}_t}(T^Y_{t-1}+T^X_{t-1})+ \mathcal{F}_t \times V^S_t  =  V^S_t$
    \State \textcolor{white}{...............}   $V^{'S}_t = T^Y_{t-1} $ 
    \State \textbf{return} $V^{'S}_t$
    \end{algorithmic}
    \end{algorithm}   

    \noindent
    Similarly, dynamics for the opposite trade, i.e. where asset $Y$ is swapped to asset $X$, can be understood using algorithm above, by simply replacing $T^X_{t-1}$ and $T^X_t$ with  $T^Y_{t-1}$ and $T^Y_t$, and correspondingly $T^Y_{t-1}$ and $T^Y_t$ with  $T^X_{t-1}$ and $T^X_t$.\\
    \\
    Finally, we ascertain the amount of the asset $Y$, that a trader can get from $L^Y$ pool, $V_t^{Y}$. We do this by converting \$S after rebalancing premium adjustments, $V_t^{'S}$, to asset $Y$. Then, the trader takes $V_t^Y$ units of of asset $Y$ out of the inventory pool $L^Y$.\\
    \\
    In conclusion, it is noteworthy that this methodology and the accounting asset, eliminate the need of pairing liquidity pools with each other.

\subsection{Rebalancing Mechanism} \label{arbvalue}
    \begin{definition}[Responsiveness Function]
        The responsiveness function regulates the size of the margin account opened by market participants and the demand for risk premia by arbitrageurs in the sLP system. Its primary purpose is to map these inputs to determine the total rebalancing premium available within the system.
    \end{definition}
    
    \noindent
    The aforementioned function enables the protocol to optimise the manner in which transitory decisions are faced by sLPs, which are of both - static and dynamic nature, whilst considering the effect of slippage, incremental inventory risk, and desired long-term stability.

\subsubsection{Rebalancing Premium Auction}
\begin{figure}[H]
      \begin{center}
          \includegraphics[width=3.3in]{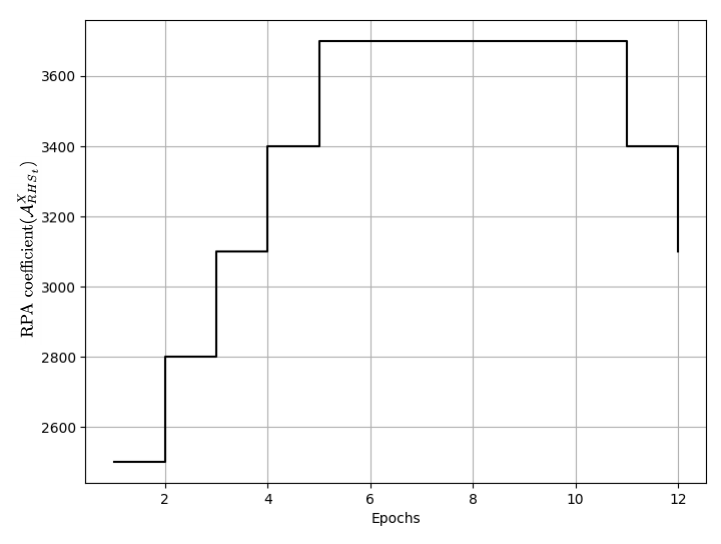}
          \caption{Rebalancing premium auction.}
          \label{fig:rpa}
      \end{center}
    \end{figure}

    DFMM utilises a series of (reverse) Dutch auctions referred to as Rebalancing Premium Auctions (RPAs) to systematically optimise the rebalancing premium. The primary objective of this optimisation is to minimise asynchronicity between the local and external markets, effectively reducing costs to the protocol.\\
    \\
    During the auction, arbitrageurs act as competing sellers, and the protocol is the buyer of the service enabling desirable rebalancing of inventory, where the bid price is the available rebalancing premium, and competitive dynamics between LPs and arbitrageurs catalyse the discovery of optimal value. Whilst at inception, the process may commence as a reverse Dutch auction, it is possible that eventually, it switches to being a Dutch auction, i.e. when there is the presence of more than a few (competing) willing providers, leading to a decline in price (offered rebalancing premium). This can be visualised, as demonstrated in Fig.\ref{fig:rpa}
    
    \noindent
    As before, we can quantify the total arbitrageable value (rebalancing premium) available at a specific aggregate trading volume level ($T_t^{X}$) of the local market, which is a function of a function:

\begin{equation}\label{rpdiscovery}
    \begin{cases}
        \mathcal{R}^{\text{X}}_{{RHS}_t}(T_t^X) = T_t^X(T_t^X + \mathcal{A}^X_{{RHS}_t}) \times \mathcal{D}^X_{{RHS}_t} & \forall T_t^X \geq 0 \\
        \mathcal{R}^{\text{X}}_{{LHS}_t}(T_t^X) = -T_t^X(-T_t^X + \mathcal{A}^X_{{LHS}_t}) \times \mathcal{D}^X_{{LHS}_t} & \forall T_t^X < 0
    \end{cases}
\end{equation}

    \begin{definition}[Utilisation Rate]
        The utilisation rate represents the state of the liquidity pools, calculated as a ratio of open inventory position to the maximum open inventory position that system can support. Let the utilisation rate for the asset $X$ in the RHS direction ($T_t^X \geq 0$) be denoted by $\mathcal{U}^X_{RHS_t}$, and similarly, $T_t^X < 0$ for LHS be denoted by $\mathcal{U}^X_{LHS_t}$, calculated as follows:
   \begin{equation}\label{rpdiscovery}
        \begin{cases}
                        \mathcal{U}^X_{RHS_t} = \frac{ \mathcal{I}^X_{LP_t}- \mathcal{I}^X_t} {min\{\mathcal{I}^X_{LP_t}, \frac{\mathcal{C}^X_{short_t}}{\varrho_{{short}_X}}\}} \\\mathcal{U}^X_{LHS_t} = \frac{ \mathcal{I}^X_t - \mathcal{I}^X_{LP_t}  } {\frac{\mathcal{C}^X_{{long}_t}}{\varrho_{{long}_X}}} 
        \end{cases}
    \end{equation}
\end{definition}
    
\begin{figure}
        \begin{center}
           \includegraphics[width=3.5in]{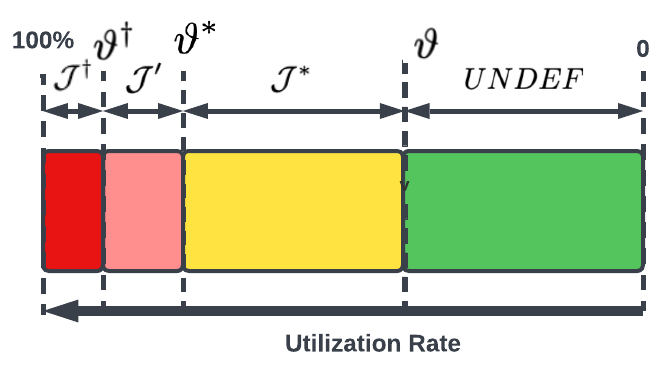}
           \caption{Rebalancing regimes, classified by utilisation rate}
           \label{fig:rebal_regimes}
        \end{center}
    \end{figure}

    \noindent
    In Fig. \ref{fig:rebal_regimes}, we introduce multiple levels of utilisation, denoted by $\vartheta$, $\vartheta^{\ast}$, and $\vartheta^{\dagger}$, following the relationship $\vartheta \leq \vartheta^{\ast} \leq \vartheta^{\dagger}$. A higher utilisation rate indicates that the system is approaching its maximum capacity to service new trades. If the system operates at a high utilisation rate, it risks potential disruptions in market operations. To ensure the sustainability of the system, there is an incentive to rebalance the inventory back to an optimal utilisation state. Therefore, the rebalancing needs are defined based on the utilisation rate of a pool. By identifying different utilisation levels, we can proactively manage the system's capacity and implement rebalancing strategies to maintain operational efficiency and stability. The first interval $[0, \vartheta]$ composes the optimum state of the utilisation rate where the system can operate optimally in terms of supporting future trades and ensuring the hedging of open inventory positions. Therefore, if the utilisation rate is within this interval, the system does not trigger the rebalancing premium auction. When the utilisation rate is within other intervals, the system aims to probe the optimality of the rebalancing function, in it's ability to bring the system back to an optimum state.
    
    \noindent
    In essence, the speed of the rebalancing process is one of the core measures of the success of the rebalancing policy, defined using rebalancing time, and contrasted against the rebalancing target, both of which are defined below. 
    
    \begin{definition}[Rebalancing Time]
        The rebalancing time ($\mathcal{T}$) represents the number of epochs (or timesteps) it takes for an arbitrageur to rebalance the protocol's inventory back to an optimal level, i.e. as defined by its utilisation rate.
    \end{definition}   

   \begin{definition}[Rebalancing Target]
        The rebalancing target is the target number of epochs(or timesteps), within which the system aims to rebalance the inventory when the system is outside of the optimum utilisation rate. This is represented by $\mathcal{J^\ast}$, $\mathcal{J}'$ and $\mathcal{J}^\dagger$ for $[\vartheta,\vartheta^\ast)$, $[\vartheta^\ast, \vartheta^\dagger)$ and $[\vartheta^\dagger, 1]$ intervals, which are system parameters.
        \begin{equation}
            \mathcal{J}^X_{RHS_{t}} =
                \begin{cases}
                     UNDEF  & \forall 0 \leq \mathcal{U}^X_{{RHS}_t} < \vartheta^X_{{RHS}_t}   \\
                    \mathcal{J^{\ast}} \leftarrow  \text{Defined by Epochs} & \forall \vartheta^X_{{RHS}_t} \leq \mathcal{U}^X_{{RHS}_t} < \vartheta^{X\ast}_{{RHS}_t}  \\
                    \mathcal{J}' \leftarrow  \text{Defined by Epochs} & \forall \vartheta^{X\dagger}_{{RHS}_t} > \mathcal{U}^X_{{RHS}_t} \geq \vartheta^{X\ast}_{{RHS}_t}\\
                    \mathcal{J}^\dagger \leftarrow  \text{Defined by Timesteps} & \forall \mathcal{U}^X_{{RHS}_t} \geq \vartheta^{X\dagger}_{{RHS}_t}.
                \end{cases}
        \end{equation}

        \noindent
        Similarly, for the system to meet the rebalancing targets following inequality must also be satisfied:
        \begin{equation}\label{rpdiscovery1}
            \mathcal{J}^X_{{RHS}_{t}} \geq \mathcal{T}^X_t.
        \end{equation}
        
    \end{definition}

    \noindent
    In essence, optimisation of rebalancing premium can be expressed as:
    \begin{equation*}
        \begin{aligned}
            & \underset{\mathcal{A}^X_t}{\text{minimise}}
            & & \mathcal{R}^{X}_{{RHS}_t}(T_t^X) \\
            & \text{s.t.}
            & & \mathcal{J}^X_{{RHS}_t} \geq \mathcal{T}^X_t.
        \end{aligned}
    \end{equation*}

    \noindent
    Thus, after initiation, if the system does not meet the rebalancing target (the inequality in Eq. \ref{rpdiscovery1}), the system triggers an iterative increase in $\mathcal{A}^X_{RHS}$ with fixed increments - $\Lambda$, which is the system parameter, influences the arbitrageable value in the system. Similarly, if the rebalancing to an optimum level is conducted much faster than the target, it can indicate an overly aggressive rebalancing premium function, triggering an incremental decrease in $\mathcal{A}^X_{RHS}$.
    
    \begin{equation}
    \mathcal{A}^X_{{RHS}_t} =
        \begin{cases}
            \mathcal{A}^X_{{RHS}_{t-1}} + \Lambda &\forall \mathcal{J}^X_{{RHS}_{t}} < \mathcal{T}^X_t\\
            \mathcal{A}^X_{{RHS}_{t-1}} - \Lambda &\forall \mathcal{J}^X_{{RHS}_{t}} > \mathcal{T}^X_t.\\
        \end{cases}
    \end{equation}
    
    \noindent
    Such an iterative process is applied by the system to enable the discovery of optimal  $\mathcal{A}^X_{{RHS}_{t}}$ in the system, and similarly the discovery of optimal arbitrageable value. The same principle can also be applied to $\mathcal{A}^X_{{LHS}_{t}}.$\\
    \\
    The auction-based change in rebalancing premia available in the internal market has a deterministic and dynamic upper bound, called Treasury Reserve($TR_t$), discussed in section \ref{protocolpnl}.
    \\
    We now define the algorithm which is used to define the auction mechanism, to ascertain the sought value:

    \begin{algorithm}[H]
    \caption{Arbitrageable value auction.}
    \label{algo:protocol2}
    \begin{algorithmic}[1]
        \State $\mathcal{A}_{{RHS}_0}^X$ $\leftarrow$   $\mathcal{A}_{{RHS}}^X$: Initialise at inception of the asset pool $X$.
        
        \State $\mathcal{A}_{{LHS}_0}^X$ $\leftarrow$ $\mathcal{A}_{{LHS}}^X$: Initialise at the inception of asset pool $X$.
        
        \State
        $\mathcal{A}_{{min}}^X$: The minimum value that $\mathcal{A}_{{RHS}_t}^X$ or $\mathcal{A}_{{LHS}_t}^X$ can accept.

        \State \textbf{if} $T^X_t > 0$   \textbf{then}    
        \State \textcolor{white}{......}\textbf{if} $ \mathcal{J}^X_{{RHS}_{t}} > \mathcal{T}^X_t$   \textbf{then}
        \State \textcolor{white}{..........}\textbf{if} $\mathcal{A}^X_{{RHS}_{t-1}} - \Lambda \geq \mathcal{A}_{{min}}^X$ \; \textbf{then} 
        \State \textcolor{white}{........................}$\mathcal{A}^X_{{RHS}_t} =\mathcal{A}^X_{{RHS}_{t-1}} - \Lambda $
        \State \textcolor{white}{..........}\textbf{else}
        \State \textcolor{white}{........................} $\mathcal{A}^X_{{RHS}_t} = \mathcal{A}_{{min}}^X$
        
        \State \textcolor{white}{......}\textbf{elif} $\mathcal{J}^X_{{RHS}_{t}} < \mathcal{T}^X_t$  \textbf{then}
        \State \textcolor{white}{..........}$ \mathcal{A}^{X'}_{{RHS}_t} = \mathcal{A}^X_{{RHS}_{t-1}} + \Lambda$
        \State  \textcolor{white}{..........}\textbf{if} $T^X_t \times (T^X_t + \mathcal{A}_{{RHS}_t}^{X'}) \times D_{{RHS}_t}^X - T^X_t \times (T^X_t + \mathcal{A}_{{RHS}_{t-1}}^{X}) \times D_{{RHS}_t}^X \leq TR_t$ \textbf{then}
        \State \textcolor{white}{....................}  $\mathcal{A}_{{RHS}_t}^{X} = \mathcal{A}_{{RHS}_t}^{X'}$ 
        \State \textcolor{white}{..........}\textbf{else}
        \State \textcolor{white}{....................} $\mathcal{A}^X_{{RHS}_t}= \frac{TR_t + T^X_t \times (T^X_t + \mathcal{A}_{{RHS}_{t-1}}^{X}) \times D_{{RHS}_t}^X-T_t^{X^2} \times D^X_{{RHS}_t}}{T^{X}_t \times \mathcal{D}^X_{{RHS}_t}}$
        \State \textbf{return} $\mathcal{A}^X_{{RHS}_t}$
    \State 
    
    \State \textbf{if} $T^X_t < 0$   \textbf{then}    
        \State \textcolor{white}{......}\textbf{if}  $\mathcal{J}^X_{{LS}_{t}} > \mathcal{T}^X_{t}$   \textbf{then}
        \State \textcolor{white}{..........}\textbf{if} $\mathcal{A}^X_{{LHS}_{t-1}} - \Lambda \geq \mathcal{A}_{{min}}^X \; \textbf{then}$ 
        \State \textcolor{white}{........................}$\mathcal{A}^X_{{LHS}_{t}} =\mathcal{A}^X_{{LHS}_{t-1}} - \Lambda $
        \State \textcolor{white}{..........}\textbf{else}
        \State \textcolor{white}{........................} $\mathcal{A}^X_{{LHS}_t} =\mathcal{A}_{{min}}^X$
        \State \textcolor{white}{......}\textbf{elif} $\mathcal{J}^X_{{LHS}_{t}} < \mathcal{T}^X_{t}$  \textbf{then}
        \State \textcolor{white}{..........}$\mathcal{A}^{X'}_{{LHS}_t} = \mathcal{A}^X_{{LHS}_{t-1}} + \Lambda$
        % \State  \textcolor{white}{..........}\textbf{if} $T^X_t \times (T^X_t + \mathcal{A}_{{LHS}_t}^{X'}) \times D_{{LHS}_t}^X -  T^X_t \times (T^X_t + \mathcal{A}_{{LHS}_{t-1}}^{X}) \times D_{{LHS}_t}^X}\leq TR_t$ {then}
        \State \textcolor{white}{..........}\textbf{if} $T^X_t \times (T^X_t + \mathcal{A}_{{LHS}_t}^{X'}) \times D_{{LHS}_t}^X -  T^X_t \times (T^X_t + \mathcal{A}_{{LHS}_{t-1}}^{X}) \times D_{{LHS}_t}^X \leq TR_t$ \textbf{then}
        \State \textcolor{white}{....................}$\mathcal{A}_{{LHS}_t}^{X} = \mathcal{A}_{{LHS}_t}^{X'}$ 
        \State \textcolor{white}{..........}\textbf{else}
        \State \textcolor{white}{....................} $\mathcal{A}^X_{{LHS}_t}= \frac{TR_t - T^X_t \times (-T^X_t + \mathcal{A}_{{LHS}_{t-1}}^{X}) \times D_{{LHS}_t}^X - T_t^{X^2}\times D_{LHS}^X}{-T^{X}_t \times \mathcal{D}^X_{{LHS}_{t}}}$
        \State \textbf{return} $\mathcal{A}^X_{{LHS}_t}$    
        \State \textbf{if} $T^X_t = 0$   \textbf{then}       
        \State \textcolor{white}{......}\textbf{pass}   
    \end{algorithmic}
    \end{algorithm}
    
    \noindent
    where, $\mathcal{A}^X_{{RHS}_0}$, $\mathcal{A}_{{min}}^X$ and $\mathcal{A}^X_{{LHS}_0}$ are system parameters.

\subsubsection{Additional sLP Incentive}
    Secondary liquidity providers essentially make markets in complex risks, which is balanced by the proportionate incentives they have to participate in the DFMM ecosystem. We discuss one of those important incentives in this subsection, which is linked to the previously defined utilisation rate ($\mathcal{U}_t$) - a measure quantifying the size of outstanding inventory using the margin vaults.\\
    \\
    Specifically, the behaviour we seek to regulate is that of depositing (or withdrawing) collateral to/from the margin vaults, in times of high (or low) utilisation rate. We incentivise sLPs to help support the long-term stability of the system, by attracting more deposits in periods of high utilisation, and encouraging withdrawals in the period of low utilisation. This helps optimise returns for all stakeholders.\\
    \\
    Mathematically, this objective is exercised through the sLP cover coefficient($\mathcal{D}^X_t$), which has the following functional form:
    
    \begin{equation}  
        \mathcal{D}_{{RHS}_t}^X = (\mathcal{D}_{{RHS}_{max}}-\mathcal{D}_{{RHS}_{min}}) \times (\frac{\mathcal{U}^X_{{RHS}_t}}{\mathcal{U}_{{RHS}_{max}}})^k + \mathcal{D}_{{RHS}_{min}}
    \end{equation}

    \noindent
    where, $\mathcal{D}_{{RHS}_{max}},\mathcal{D}_{{RHS}_{min}}$ and $\mathcal{U}_{{RHS}_{max}}$ are, respectively, the maximum and minimum values that sLP cover coefficient can accept, and $\mathcal{U}_{{RHS}_{max}}$ is the maximum utilisation rate that system deems critical.\\
    \\
    One can use the formulation above, to equivalently define $\mathcal{D}^X_{{LHS}_t}$, which is straightforward. Note, that a change in the size of the collateral vault, e.g. $\mathcal{C}^X_{{long}_t}$, can affect $\mathcal{U}_{{RHS}_t}^X$, which can be observed by changes in $\mathcal{D}^X_{{RHS}_t}$, and therefore, the aggregate rebalancing premium available in the system i.e. $\mathcal{R}^X_{{RHS}_t}(T^X_t)$. This change in the rebalancing premium is passed on to sLPs, such that an increase in the premium becomes is a cost, and a decrease in the premium is an additional source of income.\\
    \\
    Further, withdrawing collateral from the vault in times of need leads to a change from $\mathcal{C}_{{long}_{t-1}}^X$ to $\mathcal{C}_{{long}_t}^{X}$ ($\mathcal{C}_{{long}_{t-1}}^X > \mathcal{C}_{{long}_t}^{X}$), which impacts the utilisation rate ($\mathcal{U}^X_{RHS_{t-1}}<\mathcal{U}^X_{RHS_{t}}$) and the coefficient $\mathcal{D}$ from $\mathcal{D}^X_{{RHS}_{t-1}}$ to $\mathcal{D}^X_{{RHS}_t}$ ($\mathcal{D}^X_{{RHS}_{t-1}} < \mathcal{D}^X_{{RHS}_t} $). This ultimately affects an adverse change in the utilisation rate, leading to an increase in the total rebalancing premium (which we recall is the total arbitrageable value available) in the system. On the other hand, the opposite dynamics apply when the sLPs deposit collateral in times of need. It leads to a smaller utilisation rate ($U^X_{{RHS}_{t-1}} > U^X_{{RHS}_t}$) and a decrease in the sLP cover coefficient ($D^X_{RHS_{t-1}} < D^X_{RHS_t}$) and rebalancing premium. In essence, positive rebalancing premium is a  cost for sLPs, which is withdrawn from their collateral, and a negative rebalancing premium is a reward, which is deposited to their collateral.\\
    \\
    On a net-net basis, the amount payable to (or receivable from) sLPs is simply: $\mathcal{R}^{X}_{{RHS}_t}(T_{t-1}^{X})-\mathcal{R}^{X}_{{RHS}_t}(T_t^{X})$.\\
    \\
    A plot of such simulated dynamics is presented in Fig.\ref{fig:Dcoeff}:    
    \begin{figure}
        \begin{center}
            \includegraphics[width=3.5in]{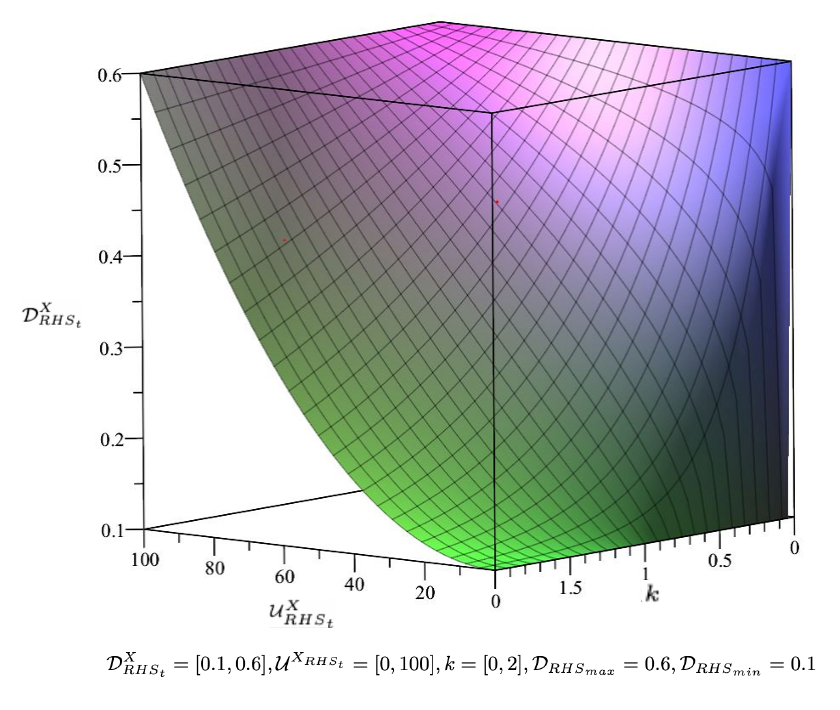}
            \caption{$\mathcal{D}$ coefficient function plot}
            \label{fig:Dcoeff}
        \end{center}
    \end{figure}
 
\subsection{Protocol P\&L}\label{protocolpnl}
    Arbitrageable value is a protocol-driven cost for the system, which needs to be sustainable in the long run. To motivate the concept, consider the following scenarios:
    
    \begin{enumerate}
        \item At time $t$, the coefficient $\mathcal{A}$ of rebalancing premium function remained constant, $\mathcal{A}^{X}_{t-1}= \mathcal{A}^{X}_{t}$, however the trading activity has altered the state of the inventory level $|T_{t-1}^{X}| < |T_{t}^{X}|$. Therefore, trades that increase the difference in states of inventory level between the epochs $t$ and $t-1$ ($T_{t-1}^{X}-T_{t}^{X}$), are executed with a worse price (for the trader), than the price available in the aggregate external markets, derived using the ELDF. This difference in execution price is a revenue for the market maker (via an increase in the available rebalancing premium), and cost for the trader. Similarly, when the rebalancing is conducted, the protocol distributes the rebalancing premium to the rebalancing agent - whether an arbitrage or trader, which is a cost for the protocol and revenue for the arbitrager. All else being equal, upon conclusion of the rebalancing process, the protocol has net zero cost.
    
        \item The coefficients of the rebalancing premium function have changed due to the rebalancing premium auction process $\mathcal{A}^X_{t-1} \neq \mathcal{A}^X_t$.
        In this scenario, there is a divergence between trader cost and rebalancing premium available in the market($\mathcal{R}^{X}_t$). We denote this divergence by $\Upsilon_t^X \in R$.  The sign of the $\Upsilon_t^X$  is dependent on the type of auction applied.\\
        \\
        If reverse Dutch auction has been applied,  $\mathcal{A}_{t-1}^X < \mathcal{A}_t^X$,  the rebalancing premium charged from traders would not be enough to cover total arbitrageable value available to incentivise moving total inventory levels back to the optimum level(positive discrepancy).

        \begin{equation}
            \Upsilon_t^X = \mathcal{R}^{{X}, \mathcal{A}_{t}}_t(T_t^{X}) - 
            \mathcal{R}^{{X}, \mathcal{A}_{t-1}}_t(T_t^{X})> 0.
        \end{equation}

        \noindent
        If Dutch auction has been applied ($\mathcal{A}_{t-1}^X > \mathcal{A}_t^X$), else the opposite will be true: 

        \begin{equation}
            \Upsilon_t^X = \mathcal{R}^{X, \mathcal{A}_{t}}_t(T_t^{X}) - \mathcal{R}_t^{X, \mathcal{A}_{t-1}}(T_t^{X}) < 0.
        \end{equation}

        \noindent
        For the protocol, the positive discrepancy is a cost and the negative discrepancy is revenue. Therefore, if no other revenue and cost sources are introduced, in the second scenario, the protocol cost can be greater (in case of reverse Dutch auction) or smaller(in case of reverse Dutch auction)  than the protocol revenue. To ensure that the system has enough capital to offset the cost, a special rebalancing fee ($\Xi_t$) has been applied in the marketplace and charged as a percentage($\xi$) on traded volume ($V_t^x$) as part of platform AMM fees.  
        
        \begin{equation}
            \Xi_t^X = V_t^X \times \mathcal{\xi}_t , \text{where} \textcolor{white}{...}  0 \leq \xi_t <1
        \end{equation}
        where $\xi_t$ is a system parameter, $\xi_t \leq \theta_t$
    \end{enumerate}

    \begin{definition}[Treasury Reserve ]
        The Treasury Reserve ($TR_t^X$) represents the maximum available reserves, correspondingly the maximum available auction-based rebalancing premium changes the system can support for all pools, without violating the protocol cost and revenue inequality, $cost \leq revenue$.
        
        \begin{equation}
            TR_t = \sum_{i=1}^{m}\sum_{n=0}^t\Xi_n^i - \sum_{i=1}^{m}\sum_{n=0}^{t} \Upsilon_n^i
        \end{equation}
    \end{definition}
    \noindent
    where the $m$ is the number of the pools in the system.\\
    \\
    In essence, the system aggregates the protocol reserves for all pools to enjoy shared resources applied to facilitate optimum re-balancing in the system.

\newpage
\subsubsection{Reward Distribution}
    The system collects AMM fee ($\theta_t$), denominated in the algorithmic accounting asset, which is used to compensate pLPs and sLPs in exchange for the services they provide to the system, represented by $\theta_{pLP}^X$ and $\theta_{sLP}^X$, respectively. These point-in-time rewards, and are eligible for withdrawal by the agents who earned it, can be quantified as below:
    \begin{equation}
        \text{Reward}_t^X =  V_t^X \times (\theta_T - \xi_t),
    \end{equation}
    
    \noindent
    where, $V_t^X$ is traded volume of asset $X$ at the timestep $t$, and $\xi^X_t$ is the portion of the collected fee that is being contributed to the treasury reserve for a specific asset.\\
    \\
    Note, that for SLP rewards, we distinguish between sLPs that contribute to the long, and short vault, by representing their fees, with an additional subscript - $\theta_{sLP_{{short}_t}}^X$ and $\theta_{sLP_{{long}_t}}^X$. Therefore, the relationship between LP rewards can be linked as follows:
    \begin{equation}
        \theta_t - \xi_t = \theta_{{pLP}_t}^X + \theta_{sLP_{{short}_t}}^X + \theta_{sLP_{{long}_t}}^X.
    \end{equation}
    
    \begin{theorem}\label{ax:first} 
        Let $\mathcal{C}^X_{{short}_t}$ be the capital allocated for the short vault of the asset $X$ at time $t$, and let $\varrho^X_{{short}_t}$ be the collateralisation ratio of the short vault. Let $\mathcal{I}^X_{t}$ be the volume of asset X available in the liquidity pool at the time $t$. The ask-side capital efficiency in the DFMM market is maximised when the following condition is satisfied:
        \begin{equation}
            \frac{\mathcal{C}^X_{{short}_t}}{\varrho^X_{{short}_t}} = \mathcal{I}^X_{t}.
        \end{equation}

        \noindent
        \textbf{Proof}: We prove the proposed theorem by contradiction.\\
        \\
        Let's start by assuming that capital efficiency is achieved when the condition stated in the theorem is not true.\\
        \\
        If $\frac{\mathcal{C}^X_{{short}_t}}{\varrho^X_{{short}_t}} < \mathcal{I}^X_{t}$, then the maximum trading volume that the system can support is limited to $\frac{\mathcal{C}^X_{{short}_t}}{\varrho^X_{{short}t}}$. This implies that a portion of the assets of the liquidity pool $\left(\mathcal{I}^X_{t} - \frac{\mathcal{C}^X_{{short}_t}}{\varrho^X_{{short}_t}}\right)$ cannot be bought due to a lack of comprehensive derivative protection.\\
        \\
        On the other hand, if $\frac{\mathcal{C}^X_{{short}_t}}{\varrho^X_{{short}_t}} > \mathcal{I}^X_{t}$, then even though there is enough derivative protection available to buy $\frac{\mathcal{C}^X_{{short}_t}}{\varrho^X_{{short}_t}}$ units of asset $X$ from the market, traders cannot buy more than $\mathcal{I}^X_{t}$ units of asset $X$ from the system because it is not available in the liquidity pool.\\
        \\
        Therefore, in both cases, a portion of the margin account $\left(\frac{\mathcal{C}^X_t}{\varrho^X} - \mathcal{I}^X_{t}\right)$ does not increase the maximum amount of asset $X$ that traders can buy from the system. This indicates that there is some residual value in liquidity pools or margin vaults that does not enhance the system's capital efficiency. Furthermore, an increase in the imbalance between the liquidity pool and the available derivative protection leads to a worse-off capital efficiency.\\
        \\
        Hence, the theorem holds true, and the bid-side capital efficiency in the DFMM market is maximised when:
        \begin{equation}
            \frac{\mathcal{C}^X_{{short}_t}}{\varrho^X_{{short}_t}} = \mathcal{I}^X_{t}.
        \end{equation}
        
    \end{theorem}

    \begin{figure}[H]
        \begin{center}
            \includegraphics[width=3.4in]{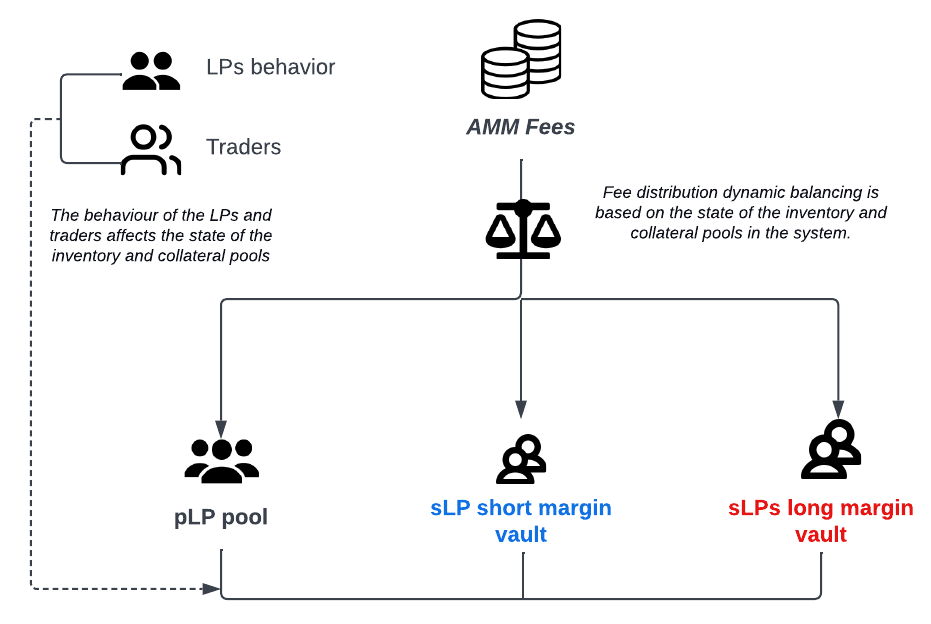}
            \caption{Dynamic balancing of an agent's incentives.}
        \end{center}
    \end{figure}

    \noindent
    On the bid-side, the system is not exposed to similar limitations of the available liquidity to trade, are users selling(depositing) liquidity to the system.  As long as there is enough derivative protection users sell the digital assets to the system and synthetically buy accounting asset from the system.\\
    \\
    At any point in time the system aims to ensure that there the system can support trades in any side of the book. More formally it can be expressed as follows:
    \begin{equation}
        \mathcal{I}^X_t = \frac{\mathcal{C}^X_{short_t}}{\varrho_{{short}_t}^X} = \frac{\mathcal{C}^X_{long_t}}{\varrho_{{long}_t}^X}.
    \end{equation}

    \noindent
    As such, DFMM seeks to discover the equilibrium price, which seeks to uniformly incentivise available protection and the size of LP pools to ensure that at any epoch the system can support balanced volume of trade at any direction. The trades that the sysem can support in the long and short directions at the begging of epoch $e$ can be calculated as follows:

    \begin{equation}
        V^{{X}_\text{bid}}_{{max}_t} =  \frac{\mathcal{C}^X_{long_e}}{\varrho_{{long}_e}^X},
    \end{equation}
    \noindent
    and,
    \begin{equation}
        V^{{X}_\text{ask}}_{{max}_t} = ,
           min({\mathcal{I}^X_e, \frac{\mathcal{C}^X_{short_e}}{\varrho_{{short}_e}^X}}).
    \end{equation}
    
    \noindent
    Since the fees of the two agents add up to the available reward in the system, finding one, leads us to be able to quantify the other.  

    \begin{algorithm}[H]
    \caption{Reward distribution.}
    \label{algo:price_adjustment}
    \begin{algorithmic}[1]
    \State \textbf{Initialiation}:
    \State $\theta^X_{LP} \gets 0.33 \times (\theta_0 - \psi^X_0)$ \Comment{Calculate some initial value}
    \State $0 \leq \mathcal{K} \leq 1$: A constant defined by decentralised governance mechanism
    \State $B \gets I^X_t + \frac{\mathcal{C}^X_{S}}{\varrho_{S}^X} + \frac{\mathcal{C}^X_{L}}{\varrho_{L}^X}$ \Comment{Calculate B using given values}
    \State $\Delta I \gets I - I^{*}$ \Comment{Calculate $\Delta I$}

    \State \textbf{Equilibrium state}:
    \State $I^X_t = \frac{\mathcal{C}^X_{S}}{\varrho_{S}^X} + \Delta I = \frac{\mathcal{C}^X_{L}}{\varrho_{L}^X} - \Delta I$

    \State \textbf{Logic}:
    \If{$\frac{I^X_t}{B} = 0.33$}
        \If{$\frac{\frac{\mathcal{C}^X_{S}}{\varrho_{S}^X} - \alpha \Delta I}{B} = 0.33$}
            \If{$\frac{\frac{\mathcal{C}^X_{L}}{\varrho_{L}^X} - \alpha \Delta I}{B} = 0.33$}
                \State $R_{SLP_L} \gets R_{SLP_S} \gets R_{PLP} \gets 0.33$
            \ElsIf{$\frac{\frac{\mathcal{C}^X_{L}}{\varrho_{L}^X} - \alpha \Delta I}{B} > 0.33$}
                \State Pay $\gamma$ from $R_{sLP_{L}}$ to $R_{sLP_{S}}$
            \Else
                \State Pay $\gamma$ from $R_{sLP_{S}}$ to $R_{sLP_{L}}$
            \EndIf
        \EndIf
    \ElsIf{$\frac{I^X_t}{B} > 0.33$}
        \State pLPs pay a fee of $\gamma$ to sLPs
        \If{$\frac{\frac{\mathcal{C}^X_{S}}{\varrho_{S}^X} - \alpha \Delta I}{B} < \frac{\frac{\mathcal{C}^X_{L}}{\varrho_{L}^X} - \alpha \Delta I}{B}$ and $R_{SLP_L} > 0.33$}
            \While{$R_{sLP_S} + 0.5\gamma > 0.33$}
                \State Pay $0.5\gamma$ from $R_{{sLP}_S}$ and $0.5\gamma$ from $R_{pLP}$
            \EndWhile
        \Else
            \State Pay $\gamma$ or ($R_{PLP}$ if $\gamma < R_{PLP}$) from $R_{PLP}$
        \EndIf
    \Else
        \State pLPs are paid a fee of $\gamma$ from sLPs
        \If{$R_{SLP_L} > 0.33 + \frac{\gamma}{2}$ and $R_{SLP_S} > 0.33 + \frac{\gamma}{2}$}
            \State $R_{SLP_L}$ and $R_{SLP_S}$ pay $\frac{\gamma}{2}$
        \ElsIf{$R_{SLP_L} > 0.33 + \frac{\gamma}{2}$ and $R_{SLP_S} < 0.33 + \frac{\gamma}{2}$}
            \State $R_{SLP_L}$ pays $\gamma$ to $R_{PLP}$
        \ElsIf{$R_{SLP_L} < 0.33 + \frac{\gamma}{2}$ and $R_{SLP_S} > 0.33 + \frac{\gamma}{2}$}
            \State $R_{SLP_S}$ pays $\gamma$ to $R_{PLP}$
        \Else
            \State $\frac{\frac{\mathcal{C}^X_{S}}{\varrho_{S}^X} - \alpha \Delta I}{B} < 0.33 + \frac{\gamma}{2}$ and $\frac{\frac{\mathcal{C}^X_{L}}{\varrho_{L}^X} - \alpha \Delta I}{B} < 0.33 + \frac{\gamma}{2}$
            \State Both $R_{SLP_L}$ and $R_{SLP_S}$ receive $\frac{\gamma}{2}$
        \EndIf
    \EndIf
\end{algorithmic}
\end{algorithm}

\section{Conclusion}
In this work, we introduced the concept of a Dynamic Function Market Maker (DFMM), designed to bridge significant gaps in the digital finance industry that we believe are a hurdle in unlocking the true potential of the DeFi industry. Our contributions encompass internal and external price aggregation mechanisms, an innovative order routing protocol, and the safeguarding of liquidity providers through a new mechanism involving sophisticated agents and a rebalancing mechanism. Additionally, we elucidate the distribution of rewards to various stakeholders within the system.\\
\\
The forthcoming version will delve into the results of our comprehensive simulation, as well as stress test outcomes. Furthermore, we will offer insightful guidance to sLP agents on appropriately pricing the digital swaptions introduced in this study.

\bibliography{main.bib}
\bibliographystyle{plain}

\end{document}